%% file: Main.tex
\documentclass[journal]{IEEEtran}

\usepackage[nolist,withpage]{acronym}
\usepackage{etoolbox}
\makeatletter
\newif\if@in@acrolist
\AtBeginEnvironment{acronym}{\@in@acrolisttrue}
\newrobustcmd{\LU}[2]{\if@in@acrolist#1\else#2\fi}

\newcommand{\ACF}[1]{{\@in@acrolisttrue\acf{#1}}}
\makeatother

\usepackage{amsmath,amsthm,amssymb}
\usepackage{mathrsfs,mathtools,mathabx}
\usepackage{bm}
\usepackage{graphicx}
\usepackage{dsfont,enumitem,stackrel}
\usepackage{color}
\usepackage{array}
\usepackage{cases}

\usepackage{algorithm,algpseudocode}
\usepackage{algpseudocode}
\usepackage{pgfplots}
\pgfplotsset{compat=1.14}
\usepackage{accents}
\usepackage{subfigure}
\usepackage{float}
\usepackage{cite}
\usepackage{xparse}
\usepackage{xcolor}
\usepackage{tikz}
\usetikzlibrary{spy}
\usetikzlibrary{arrows,decorations.pathreplacing,patterns}
\usepackage{hyperref}
\hypersetup{colorlinks,allcolors=black}
\usepackage{booktabs} 

\definecolor{brown(web)}{rgb}{0.65, 0.16, 0.16}
\definecolor{antiquewhite}{rgb}{0.98, 0.92, 0.84}
\definecolor{antiquefuchsia}{rgb}{0.57, 0.36, 0.51}
\definecolor{chestnut}{rgb}{0.8, 0.36, 0.36}
\definecolor{airforceblue}{rgb}{0.36, 0.54, 0.66}
\definecolor{cadmiumorange}{rgb}{0.93, 0.53, 0.18}
\definecolor{bleudefrance}{rgb}{0.19, 0.55, 0.91}
\definecolor{carolinablue}{rgb}{0.6, 0.73, 0.89}
\definecolor{blue(ncs)}{rgb}{0.0, 0.53, 0.74}
\definecolor{dodgerblue}{rgb}{0.12, 0.56, 1.0}
\definecolor{cssgreen}{rgb}{0.0, 0.5, 0.0}
\definecolor{cadmiumgreen}{rgb}{0.0, 0.42, 0.24}
\definecolor{cadmiumorange}{rgb}{0.93, 0.53, 0.18}
\definecolor{amaranth}{rgb}{0.9, 0.17, 0.31}
\definecolor{bluegray}{rgb}{0.4, 0.6, 0.8}
\definecolor{cadmiumgreen}{rgb}{0.0, 0.42, 0.24}
\definecolor{burntsienna}{rgb}{0.91, 0.45, 0.32}
\definecolor{ballblue}{rgb}{0.13, 0.67, 0.8}
\definecolor{dodgerblue}{rgb}{0.12, 0.56, 1.0}
\definecolor{celestialblue}{rgb}{0.29, 0.59, 0.82}
\definecolor{jazzberryjam}{rgb}{0.65, 0.04, 0.37}
\definecolor{bazaar}{rgb}{0.6, 0.47, 0.48}

\definecolor{lavendergray}{rgb}{0.77, 0.76, 0.82}
\colorlet{lavendergray2}{white!70!lavendergray}

\newtheorem{prop}{Proposition}
\newtheorem{theorem}{Theorem}

\newtheorem{remark}{Remark}	
	
\newtheorem{lem}{Lemma}	
\newtheorem{asum}{Assumption}

\theoremstyle{definition}
\newtheorem{definition}{Definition}

\def\change{\color{black}}

\graphicspath{{./Figures/}}

\begin{document}
\input{acronyms.tex}

\title{Blind Federated  Learning via Over-the-Air q-QAM}

\author{\IEEEauthorblockN{Saeed Razavikia$^\dagger$, José Mairton Barros Da Silva Júnior$^{*}$,  Carlo Fischione$^\dagger$}\\
 \IEEEauthorblockA{$^\dagger$School of Electrical Engineering and Computer Science, KTH Royal Institute of Technology, Stockholm, Sweden\\
$^*$Department of Information Technology, Uppsala University, Uppsala, Sweden\\
{Email: $^\dagger$\{sraz, carlofi\}@kth.se,}$^*$ mairton.barros@it.uu.se}
}

\maketitle

\begin{abstract}
In this work, we investigate federated edge learning over a fading multiple access channel. To alleviate the communication burden between the edge devices and the access point, we introduce a pioneering digital over-the-air computation strategy employing q-ary quadrature amplitude modulation, culminating in a low latency communication scheme. Indeed, we propose a new federated edge learning framework in which edge devices use digital modulation for over-the-air uplink transmission to the edge server while they have no access to the channel state information. Furthermore, we incorporate multiple antennas at the edge server to overcome the fading inherent in wireless communication.  We analyze the number of antennas required to mitigate the fading impact effectively. We prove a non-asymptotic upper bound for the mean squared error for the proposed federated learning with digital over-the-air uplink transmissions under both noisy and fading conditions.  Leveraging the derived upper bound, we characterize the convergence rate of the learning process of a non-convex loss function in terms of the mean square error of gradients due to the fading channel. Furthermore, we substantiate the theoretical assurances through numerical experiments concerning mean square error and the convergence efficacy of the digital federated edge learning framework. Notably, the results demonstrate that augmenting the number of antennas at the edge server and adopting higher-order modulations improve the model accuracy up to $60\%$. 
\end{abstract}


\acresetall

\section{Introduction}

In recent years, the surge in mobile device utilization coupled with advancements in sensing and machine learning technologies has facilitated the easy acquisition of mobile data to train machine learning models. Despite this, the heterogeneous and decentralized nature of these data poses a significant challenge. This is particularly evident if one wants to guarantee the optimal performance of models trained locally on individual devices when applied to data from other devices. One potential solution is centralizing data from various devices to a cloud repository for collaborative model training. However, this strategy is often impractical due to substantial communication costs, data privacy issues, and different device hardware specifications~\cite{li2020federated}.

An alternative approach gaining traction is federated learning (FL), wherein a central node aggregates local models from numerous devices to formulate a more refined global model while maintaining data privacy by keeping datasets localized on respective devices. In this scenario, the \ac{ES} oversees the progression of the global model~\cite{konevcny2016federated,azimi2022multi}. Devices compute local model updates utilizing their respective data and send them to the \ac{ES}, aggregating these updates to enhance the global model. This methodology has been further developed into \ac{FEEL}, integrating FL with edge computing to reduce communication overhead by facilitating the transmission of models to an \ac{ES} for aggregation into a cohesive model~\cite{mcmahan2017communication}.

However, this approach has limitations. As the number of devices engaged in \ac{FEEL} increases, or as the dimensions of local gradients expand, the wireless resources at the network, such as bandwidth and time, may become strained~\cite{tak2020federated}. To mitigate this, over-the-air computation (OAC) has been proposed~\cite{zhu2019broadband}. This technique concurrently allocates resources to all users, utilizing the waveform superposition of wireless signals. Furthermore, traditionally OAC employs precise analog signal precoding to facilitate the computation of specific functions over-the-air, such as arithmetic means or weighted sums, thereby enhancing efficiency in the use of the resources of the wireless channel~\cite{goldenbaum2014nomographic,hellstrom2022wireless}. However, most previous studies have focused on analog modulations for OAC and FEEL. Digital modulations would be much more favorable due to their widespread use and error-correction coding capabilities. 

Here, we aim to design a new \ac{FEEL} scheme, where the edge device uses digital modulation instead of the traditional analog modulation for communication via the OAC technique. Adopting digital modulation in OAC and retaining the benefits of the analog approach allows us to move one step towards deploying OAC with current wireless technologies and to address the \ac{FEEL} challenges, such as communication costs and data privacy, without waiting for the (perhaps economically hard to think) re-introduction of analog communications.

\subsection{Literature Review}

 Recent developments in the \ac{FEEL} framework have witnessed the exploration of advanced signal processing techniques, such as beamforming at the \ac{ES}, equipped with multiple antennas, is underway to augment the quality of the estimated signal utilized in global model updates. Notably, research has been conducted to maximize the number of devices participating in each communication round of training through the implementation of beamforming at the \ac{ES}~\cite{yang2020federated}. Additionally, innovative nonlinear estimation methods have been developed to precisely recover the sum of updates transmitted by devices, leveraging the inherent sparsity of these updates~\cite{jeon2020compressive,amiri2020federated}. Recent studies have also addressed time synchronization errors, with particular emphasis on formulating misaligned OAC into an atomic norm minimization problem to develop synchronization-free estimators~\cite{shao2021federated,razavikia2022blind,hellstrom2023optimal}. Moreover, in \cite{amiri2020federated,tegin2021blind,amiri2021blind,csahin2021distributed}, they studied \ac{FEEL} problem over wireless fading \ac{MAC}. 

In parallel, the \ac{FEEL} domain has observed substantial advancements in digital aggregation methods for wireless communication in federated learning applications. The one-bit broadband digital aggregation (OBDA) method, outlined in~\cite{zhu2020one}, aims at reducing data communication, thereby conserving bandwidth and energy. Another significant development is adopting majority vote \ac{FSK} techniques~\cite{csahin2021distributed}, which aims at harnessing modulation techniques for efficient and reliable data aggregation in wireless networks. Moreover, a phase asynchronous \ac{OFDM}-based variant of OBDA has been introduced~\cite{zhao2022broadband}, characterized by the integration of joint channel decoding and aggregation decoders, specifically designed for digital OAC applications, enhancing privacy and efficiency by eliminating the necessity for raw data sharing.

Nevertheless, these OBDA-centric methods exhibit certain limitations, predominantly confined to specific functions such as sign detection and particular machine learning training procedures, such as the signSGD problem~\cite{bernstein2018signsgd}. To widen the class of functions for digital OAC, authors in \cite{csahin2023over} use balanced number systems for computing summation functions. Despite its potential, this approach requires the allocation of unique frequencies for each quantized level, thus raising concerns regarding resource usage and spectral efficiency.  

Recently,  in \cite{razavikia2023channelcomp,razavikia2023computing, razavikia2023SumCode}, we introduced a fundamentally new communication for computation framework called ChannelComp, which allows for the computing of arbitrary functions over the \ac{MAC} with digital modulations. ChannelComp offers a spectral efficient communication scheme through digital modulation for the OAC problem, which has several computational benefits compared to analog OAC and OBDA methods and is fully compatible with existing digital communication systems.

\subsection{Our Contribution}

In this work, we propose a novel and general digital framework for \ac{FEEL}, termed ChannelCompFed, which leverages OAC over a wireless fading \ac{MAC} using q-QAM modulations. The ChannelCompFed methodology is crafted to implement digital OAC, thereby preserving the spectral efficiency intrinsic to digital systems. A significant feature of the ChannelCompFed framework is the elimination of the necessity for edge devices to be aware of the \ac{CSI}, marking a significant deviation from conventional practices in over-the-air computation over a wireless fading \ac{MAC}, where each transmitting device modulates its transmission based on the instantaneous channel state to ensure signal convergence at the same power level at the \ac{ES}.

ChannelCompFed strategically integrates multiple antennas at the \ac{ES} to mitigate the fading phenomena inherent in wireless channels, thus enhancing its operational capabilities. ChannelCompFed builds on the concept proposed in ChannelComp~\cite{razavikia2023channelcomp}, a fully digital OAC method where constellation points at the receiver are sufficiently spaced to allow the function computation. Moreover, ChannelCompFed adopts a digital coding scheme proposed in~\cite{razavikia2023SumCode}, facilitating computations over-the-air and promoting ultra-low latency communication while retaining the benefits of digital modulations. Contrary to the uncoded approaches presented in~\cite{amiri2021blind,tegin2021blind}, ChannelCompFed utilizes high-order \ac{QAM} to amplify the communication rate, thereby enhancing communication reliability and fostering superior gradient estimation, which in turn accelerates convergence during the learning process. {\change We highlight that, unlike the ChannelComp that designs new modulations diagram for computing general functions, ChannelCompFed uses a closed-form coding scheme tailored high order QAM to compute the mean function critical for FEEL. }

Furthermore, we establish the theoretical performance of the ChannelCompFed framework. We conduct a meticulous analysis to determine the optimal number of antennas at the \ac{ES} necessary to counteract the wireless channel's fading effectively.  We theoretically analyze the function computation's \ac{MSE} for noisy and fading scenarios, considering \ac{QAM} modulations of arbitrary order. To complement our theoretical insights, we prove the convergence of ChannelCompFed, thus providing a theoretical guarantee that verifies its performance efficacy. We illustrate these theoretical propositions through numerical experiments designed to evaluate ChannelCompFed's proficiency using the MNIST and CIFAR-10 datasets as benchmark.

\input{fig/Fig_Feel}

In summary, our contributions are  as follows:
\begin{itemize}

    \item {\textbf{Digital federated edge learning}: 
        We consider a digital OAC scheme, termed ChannelCompFed, for the \ac{FEEL} problems fully compatible with arbitrary order of \ac{QAM} digital communication systems. The ChannelCompFed method improves the communication rate while proving reliable communication in \ac{FEEL} network. 
    }
\item {\textbf{No \ac{CSI} information at edge device}: Employing multiple antennas at \ac{ES} allows us to design a receive beamformer for compensating the effect of fading. {\change Hence, edge devices do not require knowledge of their \ac{CSI}, and the \ac{ES} only needs to estimate the sum of the channel coefficients of all the edge devices to each antenna instead of each individual channel coefficient.}
    }
\item {\textbf{Required number of antennas}:  
        We provide the analysis for the required number of antennas, $N_r$, at the \ac{ES} to obtain an upper bound for the \ac{MSE} of the gradient over fading \ac{MAC}. In particular, we obtain a probabilistic lower bound on the number of required antennas at the \ac{ES} and show that the number of antennas has an inverse relation to the variance of the error, $\sigma^2$, i.e., $N_r = \mathcal{O}(1/\sigma^2)$. 
    }
\item {\textbf{MSE analysis and Convergence rate}:  
        We further derive the \ac{MSE} analysis of the gradient for both noisy and fading  \ac{MAC} conditions, considering \ac{QAM} modulations of arbitrary order. Utilizing the derived \ac{MSE} expressions, we establish a convergence analysis for non-convex loss functions, articulating this in terms of the \ac{MSE} or the estimated gradient over the \ac{MAC}.
    }
    \item {\textbf{Numerical experiments}: We illustrate the theoretical claims and assess the ChannelCompFed framework's performance effectiveness via numerical experiments. These experiments evaluate the algorithm's performance over fading channels and utilize the MNIST  and the CIFAR-10 datasets as a benchmark for homogeneous and heterogeneous data distributions. It is observed that increasing the number of receiver antennas and order of modulation by factors $8$ and $2$, respectively, can lead to approximately $60\%$ of improvements in the model's accuracy for classification problems.  
    }
\end{itemize}
  
\subsection{Organization of The Paper}
The remainder of this paper is organized as follows. Section~\ref{sec:system_model} presents the learning and communication models for the proposed OAC system. In Section~\ref{sec:Theoretical}, we introduce the ChannelCompFed framework and establish its performance in terms of \ac{MSE} and convergence rate for the FL problem. Then, we provide the numerical experiments to assess the performance of the proposed scheme in Section~\ref{sec:Numerical}, followed by the concluding remarks in Section~\ref{sec:Conclusion}.
\subsection{Notations}

The scalars are represented by lowercase letters, such as $x$, whereas vectors are denoted by lowercase boldface letters, $\bm{x}$. The transpose of a vector $\bm{x}$ is indicated by $\bm{x}^{\mathsf{T}}$, and its Hermitian is represented by $\bm{x}^{\mathsf{H}}$. Given two arbitrary vectors, $\bm{a}$ and $\bm{b}$, the notation $\langle \bm{a}, \bm{b}\rangle$ represents the inner product $\bm{b}^\mathsf{H}\bm{a}$. When referring to a vector $\bm{x}$, $\|\bm{x}\|$ denotes its $\ell_2$ norm. For a complex scalar $x$ belonging to the set $\mathbb{C}$, the real and imaginary parts of $x$ are represented by $\mathfrak{Re}({x})$ and  $\mathfrak{Im}({x})$, respectively.

In the context of sets, if $S$ is a set, its cardinality is denoted by $|S|$. For an integer $N$, the notation $[N]$ is used to represent the set $\{1,2,\dots, N\}$. The Normal distribution with variance $\sigma^2$ is expressed as $\mathcal{N}(0,\sigma^2)$. Additionally, $\mathcal{CN}(0,\sigma^2)$ denotes a circularly symmetric complex Normal distribution, wherein the real and imaginary terms are each distributed according to $\mathcal{N}(0,\sigma^2)$.
\section{System Model}
\label{sec:system_model}

In this section,  we provide the learning and communication models used in both the uplink and the downlink. 
  
\subsection{Learning Model}

\input{fig/FigSumCode}

In the \ac{FEEL} framework~\cite{jordan2018communication, mcmahan2017communication, konevcny2016federated}, our system model operates within a distributed learning scenario involving $K$ edge devices together with an \ac{ES} to train a shared global model, $\bm{w}$, without sharing their private data.  In this regard, each device $k$ holds a distinct local training dataset, labeled as $\mathcal{D}_k$, where $|\mathcal{D}_k|$ is the number of samples in edge device $k$.  The local loss function used by edge device $k$ is denoted by $\mathcal{L}_k(\bm{w})$, where  $\bm{w} \in \mathbb{R}^{N}$ represents the model parameters with size $N$. In the centralized learning approach, the \ac{ES} utilizes the data across all devices, i.e., $\mathcal{D}:= \mathcal{D}_1\cup \cdots \cup \mathcal{D}_K$,  and then undertakes the training of the global model by minimizing the empirical loss function in a distributed fashion:
\begin{align}
\nonumber
\bm{w}^* & =\underset{\bm{w}}{\rm arg min}~\mathcal{L}(\bm{w}) = \underset{\bm{w}}{\rm arg min}~ \frac{1}{|\mathcal{D}|} \sum\nolimits_k |\mathcal{D}_k|\mathcal{L}_k(\bm{w}), \\ & = \underset{\bm{w}}{\rm arg min}~\frac{1}{|\mathcal{D}|} \sum\nolimits_{j\in \mathcal{D}} \ell_j(\bm{w}),
\end{align}
where $\mathcal{L}(\bm{w})$ denotes the global loss function of the model vector $\bm{w}$, and $\ell_j(\bm{w})$ is the empirical loss function calculated for the $j$-th data sample of the dataset $\mathcal{D}$. 

In a bid to ensure privacy, the FL approach has been proposed. In FL, each device employs its local dataset $\mathcal{D}_k$ for conducting stochastic gradient descent (SGD) for the minimization of the local loss function $\mathcal{L}_k(\bm{w})$. Moreover, let $\bm{w}_k{(m)}\in \mathbb{R}^{N}$ and $\bm{g}_k{(m)}\in \mathbb{R}^{N}$ denote the local model parameters and gradient estimation of device $k$ at the $m$-th communication round, respectively. Specifically, in communication round $m$, device $k$ computes $\bm{g}_k{(m)}$ gradient over its local data set $\mathcal{D}_k$ as
\begin{align}
\label{eq:signlemodel}
\bm{g}_k{(m)} = \frac{1}{|\mathcal{D}_k|} \sum\nolimits_{j\in \mathcal{D}_k}\nabla \ell_j(\bm{w}_k{(m)}).
\end{align}
Next, each device's local model update $\bm{g}_k{(m)}$ is sent to the \ac{ES}. On receipt of gradient, the \ac{ES} calculates the global model of the gradient $\bm{g}{(m)}$  across all $K$ devices as:
\begin{align}
\label{eq:aggreg}
\bm{g}{(m)} = \frac{1}{K}\sum\nolimits_{k=1}^{K}\bm{g}_k{(m)}.
\end{align}
Finally, the \ac{ES} then updates the current global model by $\bm{w}{(m+1)} = \bm{w}{(m)} - \eta \bm{g}{(m)}$ where $\eta$ represents the learning rate. Afterward, the updated global model is broadcast back to the edge devices, and the process repeats until a convergence criterion, such as the maximum number of communication rounds or convergence to a local minimum, has been reached. We note that in Eq.~\eqref{eq:aggreg}, the \ac{ES} requires only the aggregation of local estimates $\bm{g}_k{(m)}$ and not individual gradients from each edge device. Therefore, we can use \ac{FEEL}, in which the per-round communication latency is independent of the number of \ac{ES}. This is a benefit of FEEL compared to the traditional multiplexing methods, such as time multiplexing and frequency multiplexing, in which the per-round communication latency and bandwidth increase linearly by the number of \ac{ES}s, respectively. The overall learning procedure is depicted in Figure~\ref{fig:federated}.

To alleviate the notation, we omit the index $m$ from $\bm{g}{(m)}$ and represent the gradient at each communication round as $\bm{g}$ because the communication protocol remains constant across the communication rounds.

\subsection{Communication Model}\label{sec:Commodel}

 We assume that the \ac{ES} is equipped with $N_r$ transmitter/receiver antennas, and every edge device has a single transmitter/receiver antenna.  In each communication round, the edge device $k$ needs to transmit its gradient $\bm{g}_k$ by the q-QAM modulated signal ${x}_{k}\in \mathbb{C}$ to the \ac{ES} over a broadband \ac{MAC}. In order to manage frequency-selective fading and inter-symbol interference, \ac{OFDM} is employed, i.e.,  fractionating the bandwidth $B$ into $N$ orthogonal subchannels~\footnote{For the ease of explanation, we assume that the number of orthogonal subchannels, $B$,  equals the size of model parameters vector $\bm{w}$, i.e., $N$. However, extension to an arbitrary number is straightforward, and one can divide the number of model parameters into the number of available communication resources, such as subchannels and time slots, and perform multiple transmissions over different subchannels and time slots~\cite{amiri2021blind,yang2022over}.}.
 
 In particular,  to perform a digital transmission, the procedure is the following: element $n$ of the gradient vector of edge device $k$, ${g}_{k}^n$, is quantized into a scalar, $\tilde{g}_{k}^n:=\mathcal{Q}_{q}(g_{k}^n)$, with $q$ possible values for $n\in [N]$, where $\mathcal{Q}_{q}(\cdot)$ is the quantizer and $q$ is the number of quantization levels. Then, the resultant vector is mapped into the digitally modulated signal ${x}_{k}^n$  using the encoder $\mathscr{E}_{q}(\cdot)$, i.e., ${x}_{k}^n = \mathscr{E}_{q}(\tilde{g}_{k}^n)$. To prevent the transmission power from influencing the learning rate~\cite{hellstrom2023retransmission}, we normalize and denormalize the modulated signals, ${x}_{k}^n$, using a pre-processing function, $\varphi_k$, and post-processing function, $\psi$,  at edge device $k$ and the \ac{ES}, respectively. Specifically, edge device $k$ first encodes ${x}_{k}^n$ into $s_{k}^n$ using a pre-processing function $\varphi_k$ as follows:
\begin{align}
    \label{eq:Sk}
      \bm{s}_{k} := \varphi_k(\bm{x}_{k}) = \frac{|D_k|\bm{x}_{k}}{\sqrt{\beta}},
\end{align}
where $\beta$ is scaling factor that is chosen to satisfy a transmit power constant for edge devices and ${\bm{x}}_{k}:= [{x}_{k,1}, \ldots, {x}_{k,N}]^{\mathsf{T}} \in \mathbb{C}^{N}$ and  ${\bm{s}}_{k}:= [{s}_{k,1}, \ldots, {s}_{k,N}]^{\mathsf{T}} \in \mathbb{C}^{N}$ and where recall that $\mathcal{D}_k$ is dataset of device $k$. For a more realistic approach, we consider that the average transmission power of every edge device is limited by a predefined positive value, $P_{\rm max}$. With the given transmitted symbols in Eq.~\eqref{eq:Sk}, we can define the power constraints as follows:
\begin{align}
    \label{eq:Powerbeta}
    \mathbb{E}\big[\|\bm{s}_k\|^2\big]  \leq P_{\rm max}, \forall~k\in [K].
\end{align}
Consequently, we have $ \beta \geq \mathbb{E}[|D_k|^2\|\bm{x}_{k}\|^2]/P_{\rm max}$, which means that $\beta \geq \max_k\{\mathbb{E}[|D_k|^2\|\bm{x}_{k}\|^2]\}/P_{\rm max}$. {\change Note that higher $\beta$ reduces the transmitted power, diminishing the SNR and impairing performance. However, the exact values of lower bounds are unavailable in practical scenarios due to the unknown magnitude of gradients beforehand. Consequently, the lower bounds must be approximated and $\beta$ chosen slightly above the lower bounds to ensure the power constraint in \eqref{eq:Powerbeta} holds.

}

In the uplink transmission step, all the nodes transmit simultaneously\footnote{All nodes and the \ac{ES} are presumed to achieve perfect synchronization. However, in cases of synchronization imperfections, analog OAC strategies~\cite{razavikia2022blind,hellstrom2023optimal}, can be implemented.} and over the same subchannel. Afterwards,  the \ac{ES} server receives the summation of all $\bm{s}_k$'s over the \ac{MAC} during one-time slot~\cite{razavikia2023computing}, i.e.,   
\begin{align}
    \label{eq:Aggnoise}
      \bm{y}^{n} = \sum\nolimits_{k=1}^{K}\bm{h}_{k}^n{s}_{k}^n+ \bm{z}^{n},\quad n \in [N], 
\end{align}
where $\bm{y}^{n}$ is physically generated by the superposition of electromagnetic waves over-the-air in the wireless channels. Moreover, the $n$-th element of $\bm{h}_{k}^n \in \mathbb{C}^{N_r}$ denotes the channel coefficient between node $k$ and the \ac{ES} server at $n$-th sub-channel and is distributed according to $\mathcal{CN}(\bm{0},\sigma_h^2\bm{I}_{N_r})$; the term $\bm{z}^{n}$ represents the \ac{AWGN}, which is distributed according to $\mathcal{CN}(\bm{0},\sigma_z^2\bm{I}_{N_r})$. Moreover, different entries of  $\bm{h}_{k}^n$ and $\bm{z}^{n}$  can be correlated while the vectors are \ac{iid} across \ac{ES} antennas and subchannels of edge devices.

Next, the \ac{ES} server applies the receiver beamforming vector $\bm{u}^{n}\in \mathbb{C}^{N_r}$ to the received signal $\bm{y}^{n}$, and it yields 
\begin{align}
    \hat{s}^{n} := \langle \bm{u}^{n}, \bm{y}^{n}\rangle =   \sum\nolimits_{k=1}^{K}\langle \bm{u}^{n}, \bm{h}_{k}^n  {s}_{k}^n \rangle+ \langle \bm{u}^{n}, \bm{z}^{n}\rangle. \label{eq:r(n)}
\end{align}
Now, we need to set the beamforming vector $\bm{u}^{n}$ to reduce the influence of the distortion caused by noise $\bm{z}^{n}$ and improve the performance of over-the-air computation. To this end, following the blind  transceiver design \cite{amiri2021blind,tegin2021blind}, we set $\bm{u}^n = \sum_{k=1}^{K}\bm{h}_{k}^n/N_r\sigma_h^2$, which yields 
\begin{align}
    \hat{s}^{n}  =  \sum_{k=1}^{K}\frac{\|\bm{h}_{k}^n\|^2}{\sigma_h^2N_r}{s}_{k}^n + \hspace{-12pt} \sum_{k,k',k\neq k'}^{K}\frac{\langle \bm{h}_{k}^n, \bm{h}_{k'}^n\rangle}{\sigma_h^2N_r} {s}_{k'}^n 
    + \hspace{-5pt} \sum_{k=1}^{K}\frac{\langle \bm{h}_{k}^n, \bm{z}^{n}\rangle}{\sigma_h^2N_r}. \label{eq:sigrec}
\end{align}
Next, let us consider that $\bm{h}_{k}^n$ and $\bm{h}_{k'}^n$ are statistically independent\footnote{ \change In case $\bm{h}_{k}^n$ and $\bm{h}_{k'}^n$ are correlated,  the second term corresponds to an interference error that does not disappear and induces bias error. Let $\sigma_{k,k'}^2$ be the covariance between channels of node $k$ and $k'$.  Regardless of ${s}_{k'}^n$, the performance of OAC is degraded proportional to the ratio of $\sigma_{k,k'}^2/ \sigma_{k}^2$.  Suppose the normalized covariance matrix of the nodes channel deviates from the identity matrix. In that case, the error subsequently increases, i.e., $ \sigma_{k,k'}^2  \approx  \sigma_{k}^2$ then the expected value of the induces bias $\Bar{e}_{\rm int}^{n}$ increases, where $\Bar{e}_{\rm int}^{n}:=\sum_{k,k', k\neq k'}^{K}\sigma_{k,k'}^2s_k^n/\sigma_h^2$ for $n\in [n]$. To better capture the statistical behavior of the error, we need to analyze the random variables in the interference terms. Regardless of the statistical dependence of $h^n_k$ and $h^n_{k'}$, Lemma~\ref{lem:Delta2} shows how fast the tails of the random variable in interference deviate from their expected values. }. Then, for a large number of antennas, $N_r \gg 1$, the signal terms approaches to~\cite{rusek2012scaling}
\begin{subequations}
    \label{eq:ashall}
\begin{align}
    \label{eq:ash2}
    \|\bm{h}_{k}^n\|^2/{N_r} & \approx \sigma_h^2, \\\label{eq:ashkhk2}
     {\langle \bm{h}_{k}^n, \bm{h}_{k'}^n\rangle}/{N_r} & \approx 0, \\
     \label{eq:ashkzk}
     {\langle \bm{h}_{k}^n, \bm{z}^{n}\rangle}/{N_r} & \approx 0,
\end{align}
\end{subequations}
where $\sigma_h^2$  is the variance of channel coefficients.
Then, by substituting  Eqs.~\eqref{eq:ashall} into Eq.~\eqref{eq:sigrec}, we obtain 
\begin{align}
    \label{eq:shatapprox}
  \hat{s}^{n} & \approx  \sum\nolimits_{k=1}^{K} {s}_{k}^n, \quad n \in [N].
\end{align}
Next, the \ac{ES} applies the post-processing function of $\psi$ on the received vector $\hat{\bm{s}}$ to denormalize it as, 
\begin{align}
\label{eq:postprocessed}
\bm{r} = \psi(\hat{\bm{s}}) = \frac{\sqrt{\beta}\hat{\bm{s}}}{\sum_{k=1}^{K}|\mathcal{D}_k|}.
\end{align}
Finally, the \ac{ES} uses the decoder $\mathscr{D}$ to obtain the global gradient descent direction, i.e., $\hat{\bm{g}} = \mathscr{D}(\bm{r})$. This communication architecture is summarized in Figure~\ref{fig:SumCompFEEL}. 
{\change Note that in \eqref{eq:shatapprox}, it is assumed that a large enough $N_r$ can counteract the effects of fading and channel noise. The details and analysis of how many antennas are needed for this approximation are discussed in Section \ref{sec:Theoretical}. }
{ \change 
\begin{remark}
    The approximation in \eqref{eq:shatapprox} is a fundamental aspect of our proposed blind transceiver scheme, which aligns with the principles underlying Massive MIMO systems~\cite{rusek2012scaling} whose feasibility in 5G technologies is promising \cite{astely2022meeting}.
\end{remark}

\begin{remark}
      The \ac{ES} needs to estimate the receiver beamforming vector $\bm{u}$. However, estimating the accumulative channel gains from all nodes towards each antenna hinders the estimation of individual channel gains $\bm{h}_{k}$. In other words, this approach reduces the channel estimation overhead, a more noticeable decrease with an increase in the number of devices $K$ or the number of the ES antennas $N_r$. Indeed, this overhead is invariant concerning $K$. Moreover, given that $\bm{h}_{k}$ follows a $\mathcal{CN}(\bm{0},\sigma_h^2\bm{I}_{N_r})$ distribution, accordingly,  $\bm{u}^{n}$  obtain a Normal distribution as $\mathcal{CN}(\bm{0},\frac{K}{N_r}\bm{I}_{N_r})$. Consequently, for sufficiently large $N_r$, i.e., $N_r \gg 1$, the receiver beamforming vector tends towards an identity vector, facilitating its estimation.
\end{remark}

}
In the next section, we propose the architecture of the encoder $\mathscr{E}_q$ and decoder $\mathscr{D}$ in detail. 
  
\section{ChannelCompFed: Encoding and Decoding}

The core concept of ChannelCompFed involves integrating ChannelComp with FEEL to enhance communication efficiency via digital modulation. To this end, we must design the digital encoders and decoders of the system model to provide a low-latency aggregation for the FEEL problem (diagram depicted in Figure \ref{fig:SumCompFEEL}).  This section explains the necessary coding and decoding schemes for its implementation.

\subsection{Encoder and Decoder}

This subsection describes how to design the encoder and decoder of ChannelCompFed.  For ease of explanation, we consider the scenario where the number of antennas at \ac{ES}, $N_r$, is large enough to compensate for the channel effects, and local datasets are assumed to have uniform sizes, i.e., $|\mathcal{D}_k| = |\mathcal{D}|/K$ for $k\in [K]$. With this consideration, Eq.~\eqref{eq:postprocessed} becomes
\begin{align}
    \label{eq:noisefree}
            {r}^{n} = \frac{1}{K}\sum\nolimits_{k=1}^{K}{x}_{k}^n, \quad n \in [N],
\end{align}
where $\psi$ and $\varphi_k$ set to be $1/{K}$ and identity functions, respectively. Note that ${x}_{k}^n$ is a digitally modulated signal, and the superposition of several digitally modulated signals does not necessarily result in valid constellation points. Indeed, the constellation points of digital modulations may overlap, so we cannot assign any function output to the points \cite{razavikia2023computing}. For example, using Gray code for the \ac{QAM} modulation leads to an incomprehensible constellation diagram~\cite{razavikia2023SumCode}. To overcome this limitation, we propose to use the non-overlapping coding idea of ChannelComp~\cite{razavikia2023channelcomp} and the specific coding scheme we proposed in \cite{razavikia2023SumCode}. 

{\change
To illustrate, let us consider a simplified computation scenario involving $K=2$ nodes, where the objective is to compute the sum function $f(s_1,s_2) = s_1 + s_2$ where $s_1,s_2 \in \{0,1,2,3\}$ using Gray coded PAM with four constellation points. i.e., $x_k = (s_k+\lfloor s_k/2\rfloor - 2\lfloor s_k/3\rfloor-3 )E_s/2$ for $k = 1,2$, where $E_s$ denotes the amplitude of the carrier signal.  Then, for the $x_1  + x_2$ the constellation points are given by $r_i = (i-4) \times  E_s$ for $i\in \{1,\ldots,7\}$. In this context, a unique challenge arises: certain constellation points, specifically $r_3, r_4,$ and $r_5$, may be required to simultaneously represent multiple sum values ($\{3,2,3\}$ and $\{2,4,6\}$). This overlap creates ambiguity, as exemplified when nodes $1$ and $2$ transmit values of $s_1 = 1$ and $s_2 = 1$, resulting in the constellation point $r_3$, and similarly, $s_1 = 0$ and $s_2 = 3$ also mapping to $r_3$. Such conflicts prevent accurate computation of the desired sum at the ES, showcasing the inherent limitation of traditional coding schemes in digital OAC.

}

To encode every value of gradient ${g}_{k}^n$ of node $k$, we use \ac{QAM} of order $q$, which represents the quantization level. Without loss of generality, we let the quantized elements of the gradient be integers, i.e.,  $\tilde{g}_{k}^n \in [0,\ldots,q-1]$ for all $n\in [N]$ and $k\in [K]$. Then, we propose the encoder $\mathscr{E}_{q}(\cdot):\mathbb{Z} \mapsto \mathbb{C}$  for  scalar $g \in \mathbb{Z}$ as follows 
\begin{align}
 \label{eq:Encode1}
    [\mathscr{E}_{q}(g)]_1& :=  g - 2^{b} \cdot \left\lfloor{g}2^{-b}\right\rfloor + \frac{1-2^{b}}{2}, \\
    \label{eq:Encode2}
   [\mathscr{E}_{q}(g)]_2& = \Big\lfloor {g}2^{-b} \Big\rfloor + \frac{1-2^{b}}{2},
\end{align}
where  $b = 0.5\log_2{(q)}$ and  $[\mathscr{E}_{q}(g)]_1$ is equivalent to the \mbox{modulo} operation, and $[\mathscr{E}_{q}(g)]_2$ is the same as the floor division operation. Then, node $k$ transmits ${x}_{k}^n = [\mathscr{E}_{q}(g)]_1 + {j}[\mathscr{E}_{q}(g)]_2  \in \mathbb{C}$ for $n\in [N]$ over the \ac{MAC}. Conversely, at the receiver, to decode the average global stochastic gradient, i.e., ${g}$, we propose the decoder $\mathscr{D}_{q}: \mathbb{C} \mapsto \mathbb{Z}$ as 
\begin{align}
    \nonumber
\mathscr{D}_{q}(x):= & \mathcal{R} (\mathfrak{Re}{(x)}) + 2^{b} \cdot \left(\mathcal{R}(\mathfrak{Im}{(x)}) + \tfrac{2^{b}-1}{2}\right) \\ \label{eq:decode}
&+ \frac{2^{b}-1}{2},
\end{align}
where $x \in \mathbb{C}$ and $\mathcal{R}(\cdot)$ is the round up to half function, i.e., 
\begin{align}
   \mathcal{R}(z) =  \lceil z + 0.5\rceil - 0.5.
\end{align}
Note that this decoding function works assuming that $g$ is an integer; otherwise, the output becomes the quantized version of input $g$, i.e., $\Tilde{g} \in \{0,\ldots, q-1\}$. The decoding and encoding functions $\mathscr{D}_{q}$ and $\mathscr{E}_{q}$ satisfy the following property. 
\begin{prop}\label{prop:EncDec}
    Let $\tilde{g}_k\in \{0,1,\ldots, q-1\}$ be an integer value for $k\in [K]$ and $\tilde{g} = \sum_{k=1}^K\tilde{g}_k/K \in \{0,1/K,\ldots,q-1-1/K,  q-1\}$ be the average value. Then, the decoding and encoding functions $\mathscr{D}_{q}$ and $\mathscr{E}_{q}$ are identity operators with respect to $\tilde{g}$, i.e., 
    \begin{align}
         \tilde{g} =  \mathscr{D}_{q}\Big(\mathscr{E}_{q}(\Tilde{g})\Big) = \frac{1}{K}\mathscr{D}_{q}\Big(\sum\nolimits_{k=1}^K \mathscr{E}_{q}(\tilde{g}_{k})\Big).
    \end{align}
\end{prop}
\begin{proof}
    See Appendix \ref{sec:ProofDeCode}. 
\end{proof}

By applying Proposition~\ref{prop:EncDec} on each element of the gradient vector $\bm{g}$, we obtain the quantized version of the global model $\Tilde{\bm{g}}:= \sum_{k=1}^K\Tilde{\bm{g}}_k/K$ as follows:
\begin{align}
   \hat{\bm{g}}&  =  \frac{1}{K}\mathscr{D}_{q}\Big(\sum\nolimits_{k=1}^K \mathscr{E}_{q}(\bm{g}_{k})\Big)= \Tilde{\bm{g}}. \label{eq:estimtedg}
\end{align}
Therefore, we can successfully compute the average over the \ac{MAC} with the aforementioned coding scheme. 
{ \change
 \begin{remark}
  The encoding and decoding framework, characterized by equations \eqref{eq:Encode1} to \eqref{eq:decode}, offer several advantages over existing schemes. It ensures a unique and non-overlapping representation of aggregated signals, which is critical for accurate OAC computation. The scheme is robust to signal distortion and noise, enhancing the reliability of gradient aggregation. It facilitates efficient utilization of the communication channel, optimizing the bandwidth and energy consumption, i.e., there is no need to use orthogonal communication resources to avoid destructive overlaps. 
 \end{remark}
}
{ \change
\begin{remark}
 \label{rem:Quantization}
    Note that due to the quantization of the parameter model at edge device $k$,  $\bm{g}_k$, the global model $\bm{g}$ becomes also quantized with a finer grid $q' = q\times K$. In the case of a massive number of devices $K\gg 1$, the effect of quantization with $q'$ is negligible compared to the local model grid $q$. This phenomenon is also reflected in Proposition~\ref{Pr:MSE}, which shows that the quantization error decreases by a factor of $1/K$. As a result, quantized SGD performs similarly to the vanilla SGD algorithm.  
\end{remark} 
 
The proposed coding scheme promotes high-rate communication via over-the-air computation employing high-order QAM modulation, achieving a more effective constellation diagram than its analog equivalent. Note that the computational complexity of ChannelCompFed is almost identical to the analog scheme. Indeed, ChannelCompFed uses the SumComp coding scheme for the communication scheme, whose complexity is analytically studied as detailed in \cite{razavikia2023SumCode}. In the next subsection, the performance of ChannelCompFed with the proposed coding scheme is analyzed.

}
  
\section{Theoretical Convergence Analysis}\label{sec:Theoretical}

Here, we present the convergence analysis of \ac{FEEL} problem using the ChannelCompFed framework for both the noisy and fading \ac{MAC}. We divide this section into two parts; first, we analyze the \ac{MSE} error on the gradient over the noisy \ac{MAC} in the following subsection. Then, we provide a probabilistic upper bound on the gradient estimation error in the presence of fading as a function of the number of required antennas.

\subsection{MSE Analysis}
To provide the convergence rate of the proposed ChannelCompFed scheme, we need an upper bound on the \ac{MSE} of induced errors over the \ac{MAC}. In Proposition~\ref{Pr:MSE},  we propose an upper bound for describing the \ac{MSE} error in the \ac{AWGN} channel where the beamforming vectors have fully compensated the fading effect.   In particular, the received signal at \ac{ES} server is only contaminated by the channel noise, i.e., 
\begin{align}
    \label{eq:Aggnoisycase}
    \hat{s}^n = \sum\nolimits_{k=1}^Ks_k^n + \tilde{z}^n,\quad  n \in [N],
\end{align}
where the variance of the noise $\tilde{z}^n$ is reduced by the number of antenna, i.e., $\sigma_z^2/N_r$. In \eqref{eq:Aggnoisycase}, we remove the fading effect as reflected in \eqref{eq:Aggnoise}. 
\begin{prop}
    \label{Pr:MSE}
    Consider a communication network with $K$ nodes where each device uses the encoder $\mathscr{E}$ to compute the averaging  $\bm{g} = \sum_{k}\bm{g}_k/K$, where $\bm{g}_k$ is the local gradient of the $k$-th edge device over the noisy \ac{MAC} with $\mathcal{CN}(\bm{0}, \sigma_z^2\bm{I}_{N_r})$. Assume gradient values are generated uniformly at random with maximum absolute value $\Delta_g$, i.e., ${g}_k^n\sim \mathcal{U}(-\Delta_g,\Delta_g)$ for $n \in [N]$. Then, the \ac{MSE} of the computation errors using  \ac{QAM} modulation of order of $q = 2^{2b}$ (perfect square number), where $2b$ is the number of bits, is given by 
    \begin{align}
     \label{eq:MSEGradiant}
           \mathbb{E}[  \|\bm{g}- \hat{\bm{g}}\|^2] \leq \sigma_{\rm AWGN}^2  + \sigma_q^2,
    \end{align}
    where
    \begin{align}
    \label{eq:awgn}
        \sigma_{\rm AWGN}^2 & := \frac{N}{K^2}(1+q)e_{r},  \\
        \sigma_q^2 & :=  \frac{N\Delta_g^2}{3Kq^2},
    \end{align}
    in which $\hat{\bm{g}}$ is the estimated global gradient received in Eq.~\eqref{eq:estimtedg}, and $e_{r}$ is defined as
    \begin{align}
        \label{eq:xerror}
         e_{r} &=  2  \sum\nolimits_{\ell=1}^{2^b-1}\Big(2\ell-1 \frac{3\ell-3\ell^2 -1}{2^b}\Big) Q\Big(\frac{(2\ell-1)\sqrt{N_r}}{2\sigma_z}\Big),
    \end{align}
    and $Q(\cdot)$ denotes the Q-function.
\end{prop}
\begin{proof}
    The proof is provided in Appendix \ref{sec:ProofMSE}.  
\end{proof}

 Proposition~\ref{Pr:MSE} shows that increasing the modulation order while improving the variance of the quantization $\sigma_q^2$ can add noise term to  $e_r$. Also, increasing the number of antennas $N_r$ reduces the error by dampening the variance of the noise.

\subsection{Number of Antennas for Convergence}

We recall that in Section~\ref{sec:Commodel}, the received signal at the \ac{ES} experiences fading while the \ac{ES} can cancel the effects of fading using a large number of antenna $N_r$. From the law of large numbers, we know that when $N_r \rightarrow \infty$, the interference and the noise terms in Eq.~\eqref{eq:ashall} asymptotically vanishes. Consequently,  the estimated value of the gradient  $\hat{\bm{g}}$ approaches the actual value, i.e.,  $\bm{g}$. Therefore, we here provide a probabilistic lower bound for the number of antennas $N_r$ in the $\hat{\bm{g}}$ estimation error. In particular, we prove in Theorem~\ref{th:NRerror} that for large values of  $N_r$ and with high probability, the estimation error of $\hat{\bm{g}}$ is bounded. 

\begin{theorem}\label{th:NRerror}

Consider a communication network with $K$ edge devices and an \ac{ES} as the server equipped with $N_r$ antennas. Let the complex signal ${s}_k^{n} \in \mathbb{C}^{N}$ be the transmitted signal by the edge device $k$ over the fading channel with coefficients $\bm{h}_k^{n}$, and let $\bm{y}^n$ be the received signal at \ac{ES} for the $n$-th sub-channel. Then, the absolute difference between ${s}^n$ and its estimated value, $\hat{s}^n$,  as well as the expected value of the difference, are bounded, i.e.,  
\begin{align}
    \label{eq:epsilonupp}
    |\hat{s}^n - s^n| & \leq \epsilon, \\ \label{eq:Expecepsilonupp}
    \mathbb{E}[|\hat{s}^n - s^n|] & \leq  \frac{4K\gamma_n}{\sqrt{N_r}c_n} (\sqrt{\pi} +\ln{(6K)}),
\end{align}
where $c_n =  1/\gamma_n + \sigma_h/{\sigma_z}$ and $\gamma_n$ is a positive constant, and $N_r$ fulfills the following lower bound: 
\begin{align}
\label{eq:UpperboundNR}
     N_r \geq  \frac{8\gamma_n^2K^2}{\epsilon^2c_n^2} \ln{\Big(\frac{6K}{\delta}\Big)},
\end{align}
    with probability no less than $1-\delta$.
\end{theorem}
\begin{proof}
    The proof is provided in Appendix~\ref{sec:proofNRth}.
\end{proof}

\begin{prop}\label{cor:norm2}
    Let $\bm{g}\in \mathbb{C}^{N}$ be the global gradient averaged by the \ac{ES}, and $\hat{\bm{g}}$ be the estimated gradient of $\bm{g}:=\sum_{k}\bm{g}_k/K$, where ${g}_k^n\sim \mathcal{U}(-\Delta_g,\Delta_g)$ for $n \in [N]$.  Then, with probability no less than $1-\delta$, the error of the estimated gradient, $\hat{\bm{g}}$, as well as the \ac{MSE} of $\hat{\bm{g}}$, are bounded by scalar $\sigma_{\rm fad}^2$, i.e., 
    \begin{align}
        \label{eq:ErrorProb}
        \|\tilde{\bm{g}}- \hat{\bm{g}}\|& \leq \epsilon_{\rm fad}, \\\label{eq:ExpectErrorProb}
          \mathbb{E}[\|\bm{g}-\hat{\bm{g}}\|^2] &  \leq \sigma_{\rm fad}^2 + \sigma_q^2,
    \end{align}
    where 
    \begin{align}
        \label{eq:sigma_fed}
           \sigma_{\rm fad}^2 := 16\frac{N\gamma_{\rm max}^2 q}{N_rc_{\rm min}^2} (\pi +2\ln{(6K)}^2),
    \end{align}
    if the number of antennas, $N_r$, is greater than 
    \begin{align}
    \label{eq:AntennaGradient}
    N_r \geq  \frac{16\gamma_{\rm max}^2 N q}{\epsilon_{\rm fad}^2c_{\rm min}^2} \ln{\Big(\frac{6K}{\delta}\Big)},      
    \end{align}
     where $\gamma_{\rm max}: = \max_{n}\gamma_n$, $c_{\rm min}: = \min_{n}c_n$, and $q$ is the order of modulations. 
\end{prop}
\begin{proof}
     The proof is provided in Appendix~\ref{sec:APMSE}.  
\end{proof}

As reflected in Proposition~\ref{cor:norm2}, the number of antennas $N_r$, has an inverse relation with respect to the variance of error, $\sigma_{\rm fad}^2$. Contrary to $\sigma_{\rm AWGN}^2$,  $\sigma_{\rm fad}^2$ does not improve with the number of edge devices in the network. This is due to the choice of $\bm{u}^n$ as the sum of random channel coefficients of $K$ edge devices. Due to the variance of channel coefficients, $\sigma_h^2$, the added randomness makes it challenging to concentrate on the mean because we are averaging over the number of antennas, not the number of edge devices.

\subsection{Convergence Analysis}

In this subsection, we prove the convergence analysis of the ChannelCompFed in terms of the optimality gap for the noisy and fading channels. First, we need to present our assumptions on the loss function and the gradients, which are standard in the literature~\cite{csahin2023over,bernstein2018signsgd,zhu2020one}. 

{ \change 
\begin{asum}\label{as:baised}(\textbf{Unbiased average local stochastic gradients}) The average stochastic gradient vector $\sum_k\hat{\bm{g}}_k(m)/K$ is an unbiased estimate
of the global gradient vector $\bm{g}(m)$, i.e., $\mathbb{E}[\sum_k\hat{\bm{g}}_k(m)/K] = \bm{g}(m)$. 
\end{asum}
}
\begin{asum}\label{as:smooth}(\textbf{Smoothness}) The gradient of the loss function $\mathcal{L}(\bm{w})$ is differentiable and Lipschitz continuous with a non-negative constant $L$ on $\mathbb{R}^{+}$. This means that for any two vectors $\bm{w}$ and $\bm{v}$, the following inequalities are satisfied and equivalent:
\begin{align}
    |\mathcal{L}(\bm{w}) - \mathcal{L}(\bm{v})- \langle \nabla \mathcal{L}(\bm{w}), \bm{w}- \bm{v}\rangle| & \leq \frac{L}{2}\|\bm{w}-\bm{v}\|, \\
    \|\nabla \mathcal{L}(\bm{w}) - \nabla \mathcal{L}(\bm{v})\|& \leq {L}\|\bm{w}-\bm{v}\|.
\end{align}
\end{asum}
\begin{asum}\label{as:GD}(\textbf{Gradient Divergence}) The local gradient estimate $\bm{g}_k$ at edge device $k$ is an unbiased estimate of the global gradients $\bm{g}$ with bounded variance as
\begin{align}
    \label{eq:defgamma}
    \mathbb{E}\big[\|\bm{g}_k-\bm{g}\|^2\big]\leq \theta_k, \quad k\in [K].
\end{align}
\end{asum}
\begin{asum}\label{as:UQ}(\textbf{Unbiased quantization}) Let us denote the aggregated gradient with OAC as $\tilde{\bm{g}} = \bm{g} + \bm{e}_q,$ where $\bm{e}_q$ is the error due to the quantization. Then, the following upper bound on $\bm{e}_q$ is satisfied:
\begin{align}
    \mathbb{E}[\bm{e}_q] = \bm{0}_q, \quad \sigma_{q}^2 :=  N \frac{\Delta_g^2}{q^2K}  \geq \mathbb{E}[\|\bm{e}_q\|^2],
\end{align}
where $\Delta_g$ denotes the maximum value for a gradient element across the whole training procedure, i.e., $\Delta_g: = \max_{k,n,m}|g_k^{n}(m)|$. 
\end{asum}

Finally, using Assumptions~\ref{as:baised}-\ref{as:UQ} and the obtained upper bound on the estimated gradient from Theorem~\ref{th:NRerror} and Proposition~\ref{Pr:MSE}, we can prove the convergence of the ChannelCompFed scheme for the class of non-convex cost functions as follows.

\begin{prop}\label{prop:ConvAWGN}
    Consider a learning rate $\eta$. Moreover, consider that Assumptions~\ref{as:baised}-\ref{as:UQ}, Theorem~\ref{th:NRerror}, and Proposition~\ref{Pr:MSE} hold. Then,
    , the proposed distributed training scheme, termed ChannelCompFed, convergences to a stationary point after performing $T$ communication rounds, with the following convergence rate  
\begin{align}
\nonumber
    \mathbb{E}\Big[& \frac{1}{T}\sum\nolimits_{m=1}^{T}\|\bm{g}{(m)}\|^2\Big]  \leq \frac{1}{T\eta(1-\tfrac{\eta L}{2})}\Big(\mathcal{L}(\bm{w}{(1)})-\mathcal{L}^{*}\Big) \\ &+ \frac{\tfrac{\eta L}{2}}{1-\tfrac{\eta L}{2}}\Big((\sigma_{\rm ch}^2+\sigma_{q}^2)+\bar{\theta}\Big), \label{eq:Converg}
\end{align}
where $\bar{\theta} = \sum_{k=1}^K\theta_k/K$ with $\theta_k$ defined in Eq.~\eqref{eq:defgamma} and $\sigma_{\rm ch}^2: = \mathbb{E}[\|\bm{g}(m)-\hat{\bm{g}}(m)\|^2]$.  
\end{prop}
\begin{proof}
    The proof is provided in Appendix~\ref{Ap:ConvAWGN}. 
\end{proof}

\begin{remark}
    Note that $\sigma_{\rm ch}^2$ is the error of gradient estimation caused by the non-idealities of the communication channel. For the case of the noisy and fading \ac{MAC},  $\sigma_{\rm ch}^2$ can be upper bounded by $\sigma_{\rm AWGN}^2$ from Eq.~\eqref{eq:awgn}  and $\sigma_{\rm fad}^2$,  from Eq.~\eqref{eq:sigma_fed}, respectively. 
\end{remark}

{\change
\begin{remark}
     In practical wireless systems, electromagnetic interference follows heavy-tailed distributions \cite{clavier2020experimental} and \cite{middleton1977statistical}.  The symmetric $\alpha$-stable distribution can encapsulate the statistical behavior of such interference. The effect of such distribution in distributed learning through OAC has been studied in \cite{yang2021revisiting}. Indeed, it can be shown that the convergence trajectory is reduced by the order of $\mathcal{O}(1/T^{\alpha - 1})$, where the parameter $\alpha$ is the tail index. An inversely proportional relationship exists between $\alpha$ and the heaviness of the interference distribution's tail, with a small $\alpha$ resulting in a slow convergence rate for the learning algorithm. However, it is noteworthy that despite the corruption of the aggregated gradient by random channel fading and heavy-tailed interference with potentially infinite variance, gradient descent model training employing a decreasing learning rate can converge to the global optimum regardless of the loss function or the algorithm~\cite{yang2021revisiting}. 
\end{remark}

}

In the next section, we evaluate ChannelCompFed with respect to the impact of the error on the MSE for different numbers of antennas and evaluate the learning performance of the ChannelCompFed scheme for distributed learning.

\section{Numerical Experiments}\label{sec:Numerical}
\input{fig/Fig_Concenrtare}
\input{fig/Fig_Gradient}
\input{fig/Fig_MNIST}
In this section, we initially conduct a numerical examination of the \ac{MSE} analysis. Subsequently, we assess the {\change{efficiency }} of the digital ChannelCompFed in terms of learning accuracy and {\change{latency}} over the number of antennas and various modulation orders for homogeneous and heterogeneous data distribution.  
\subsection{MSE Analysis}

First, we analyze the performance of \ac{ES} in removing the fading effects for estimating the summation value $s^n= \sum_{k=1}^Ks_k^n$ in Eq.~\eqref{eq:shatapprox}, which results in the estimation of gradient $\hat{\bm{g}}$ over the fading \ac{MAC} with $K=200$ nodes. The transmitted signal by device $k$, $s_k^n$, is generated uniformly at random between $0$ and $1$, i.e.,  $s_k^n \sim \mathcal{U}[0,1]$ for $k\in [K]$ and any sub-channel $n \in [N]$. Then, the channel coefficient is generated randomly with Normal distribution, i.e., $\bm{h}_k\sim \mathcal{CN}(\bm{0},\sigma_{h}^2\bm{I}_{N_r})$ with $\sigma_{h}^2=1$ for $k\in  [K]$, aligning with standard assumptions in the literature, such as~\cite{amiri2021blind}. We also generate the channel's noise with the same distribution, i.e., $\mathcal{CN}(\bm{0},\bm{I}_{N_r})$. We repeat the experiment over $100$ Monte Carlo trials.

In Figure~\ref{fig:ConcentrateS}, we show the estimation performance of Eq.~\eqref{eq:shatapprox} in terms of the \ac{MSE} error for different number of antennas at the \ac{ES}. We also depict both the proposed probabilistic upper bound on the value of the error $|\hat{s}^n-s^n|$ and \ac{MSE}, from Theorem~\ref{th:NRerror}, for comparison with the outage of empirical error values. Furthermore, we show the \ac{MSE} of estimating the global gradient over the noisy \ac{MAC} using ChannelCompFed encoder and decoders in \eqref{eq:estimtedg} and analytical results from Proposition~\ref{Pr:MSE} for different numbers of antennas, $N_r$.  Note that the experiments are done for $100$ Monte Carlo trials while only $10$ trials are depicted in Figure~\ref{fig:ConcentrateS}.

Figure~\ref{fig:ConcentrateS(a)} shows the empirical \ac{MSE}, the analytical upper bound of \ac{MSE}, and the analytical upper bound on the absolute error $|\hat{s}^n-s^n|$ with the down-side triangle marker, dashed line,  and up-side triangle marker, respectively. Also, the circles in Figure~\ref{fig:ConcentrateS(a)} represent the absolute value of the estimation error from a single realization of the instantaneous Monte Carlo trial.

 Similarly, we show the empirical \ac{MSE}, the analytical upper bound of \ac{MSE}, and the error for the estimated global gradient $\hat{\bm{g}}$  with $N=100$ and $K=20$ in Figure~\ref{fig:ConcentrateS(b)}. We note that the upper bound on the \ac{MSE} is well approximated to the true value, which means it predicts the outage of the empirical error well. Moreover, by increasing the number of antennas $N_r$, the \ac{MSE} decreases, which is expected due to Theorem~\ref{th:NRerror}. This observation is consistent with the theoretical approximation of the  \ac{MSE} in Proposition~\ref{cor:norm2}.  

{ \change
 Finally,  we compute the average performance under the quantization gradient and performance under the true gradient in  \eqref{eq:MSEGradiant}. To this end, we generated edge device gradients  $\bm{g}_k$ uniformly at random from $\mathcal{U}[0,32]$ for two cases of $K=50$ and $K=400$ edge devices. Then, the gradient's continuous values are quantized by $q=64$ levels and transmitted by QAM $64$ over the noisy MAC in \eqref{eq:Aggnoisycase}. Also, the number of parameters is set to be $N=100$. This experiment is repeated by changing the distribution of gradients to  $\mathcal{U}[0,64]$ to see the effects of $\Delta_g$ on the quantization error. 

 Figure \ref{fig:Gradient}  shows  Monte Carlo numerical evaluation of the average gradient estimation in \eqref{eq:MSEGradiant} with true gradient for $100$ trials versus the analytical results from Proposition~\ref{Pr:MSE}. As noted by Remark \ref{rem:Quantization}, by increasing the number of edge devices $K$, the quantization error decreases, and ChannelCompFed performs almost equally to vanilla SGD. Moreover, the computation error is positively related to the maximum value for a
gradient element, and increasing the maximum value of the gradient results in a higher error. This observation is consistent with the theoretical approximation of the  \ac{MSE} in Proposition~\ref{Pr:MSE}. 
}

\subsection{Federated Edge Learning}

\input{fig/Fig_IIDMNIST}

We assess the performance of the proposed digital blind \ac{FEEL} scheme, i.e., edge devices have no access to the \ac{CSI}, in terms of the learning of accuracy with respect to the test dataset. We analyze two scenarios: first, including the impact of the number of antennas on the convergence in both homogeneous and
heterogeneous data distributions for the \ac{FEEL}; second, to show the level of improvement for various orders of digital modulations.  The machine learning task is image classification for the MNIST~\cite{lecun1998mnist} dataset and CIFAR-10~\cite{krizhevsky2009learning}. We use the ADAM optimizer~\cite{kingma2014adam} to train \ac{CNN}, and the size of the local mini-batch is set to be $128$. In the training process, edge devices run $3$ epochs per communication round for $T = 100$ global communication rounds.

The \ac{FEEL} network structure follows the communication model presented in Section~\ref{sec:Commodel} with $K=20$ edge devices.  The channel coefficients and noise for all sub-channels are generated according to $\bm{h}_k^{n}\!\sim\!\mathcal{CN}(\bm{0},\sigma_h^2\bm{I}_N)$ and $\bm{z}^{n}\sim \mathcal{CN}(\bm{0},\sigma_z^2\bm{I}_N)$. 

For the homogeneous data distribution, the training dataset randomly distributes into $K$ disjoint local datasets, each assigned to an edge device. The local datasets consist of samples with the same labels assigned to each edge device to model heterogeneous data distributions.

The architecture is designed for image classification tasks and comprises two 2D convolutional layers with $20$ and $40$ filters, both utilizing a \(7 \times 7\) kernel and ReLU activation. These are followed by \(2 \times 2\) max-pooling layers. Subsequently, we have a fully connected layer with $2560$ units and ReLU activation. A dropout layer with a rate of $0.2$ is included for regularization. The output layer is fully connected with $10$ units and employs a softmax activation function. The model employs Sparse Categorical Cross-Entropy as the loss function. The total number of parameters is $N = 5,086,010$. {\change Note that this architecture is used for both datasets, and the only difference is the size of the input layer, which is $28\times 28$ for the MNIST and $32\times 32$ for the  CIFAR-10 datasets, respectively. }

The performance of the proposed digital scheme is compared to the baseline (error-free) and blind \ac{FEEL} scheme proposed in \cite{amiri2021blind}. In the error-free scheme, there is no noise or fading; therefore, the \ac{ES} receives perfect copies of the model updates. Moreover, the analog \ac{FEEL} scheme follows similar communication in Eq.~\eqref{eq:sigrec} while using analog amplitude modulation instead of the SumComp in our scheme.

In Figure~\ref{fig:MNIST}, we analyze the impact of the number of antennas on the performance of the ChanneCompFed for different orders of \ac{QAM} and various levels of noise and fading. Figure~\ref{fig:MNIST(a)}  shows the test accuracy over the communication rounds for various numbers of antennas at the \ac{ES}, $N_r=\{10,20,100\}$, and \ac{QAM} orders, $q = \{64, 256\}$.  Note that increasing the number of antennas makes the convergence faster and helps the final models to reach higher accuracy, as expected according to Proposition~\ref{prop:ConvAWGN}.  Furthermore, increasing the order of modulations from $64$ to $256$ improves the learning performance, resulting in higher model accuracy after the total communication rounds. 
Figure~\ref{fig:MNIST(b)} shows a similar curve for high noise scenarios, now with $\sigma_z^2 = 10$  and number of antennas selected from $N_r=\{10,100,800\}$.  Due to the high variance of the channel noise, the \ac{ES} needs to employ both more number of antennas and a higher order of modulation to reduce the effect of fading and noise. This behavior is noticeable because the \ac{ES} fails to obtain a good accuracy (at least $60\%$) for  $N_r=10$ or $q=64$.  In this case, to compensate for the effects of fading and high noise variance, we require employing more antennas ($N_r\geq 800$)  and \ac{QAM} with $q \geq 256$.

The experiments reveal two distinct findings. In the first scenario, for noise scenario with variance on the level of the fading channels, ChanneCompFed works well provided that \ac{ES} employs either a high number of antennas, $N_r$, or high order modulation $q$. However, for a high noise scenario, with $10$ times the variance of the fading channel, there is a necessity for high-order modulations coupled with a substantial number of antennas. This event is correctly captured by Theorem~\ref{th:NRerror} and Proposition~\ref{cor:norm2} in terms of scalar $c_{n}$.

{\change 
For the next experiment presented in Figure~\ref{fig:IIDMNIST}, we use the same learning procedure from Figure~\ref{fig:MNIST} but now for homogeneous dataset scenarios for MNIST and CIFAR-10. We compare the performance of ChannelCompFed with $q=256$ to the analog scheme \cite{amiri2021blind}. We make this comparison to test the accuracy of \ac{FEEL} over a network with the same number of devices as in Figure~\ref{fig:MNIST}, but with different numbers of antennas, $N_r=\{100,400\}$ at the \ac{ES}. Figure~\ref{fig:IDDlearning} shows that the ChannelCompFed and analog \ac{FEEL} perform well for the homogeneous with $N_r= 400$ antennas for both datasets. However,  their performance degrades for the case $N_r = 100$  due to the high fading and noise variances.  By increasing the number of antennas $N_r$ for the \ac{ES}, the performance of both analog and ChannelCompFed asymptotically reaches the error-free baseline. Figure~\ref{fig:IIDCIFAIR} shows a similar scenario for $K=10$ edge devices with different channel noise and fading variances where $\sigma_z^2 = \sigma_h^2 = 2$ for the CIFAR-10 dataset. While both the ChannelCompFed and analog \ac{FEEL} perform similarly for $N_r= 400$ and $N_r =100$, ChannelCompFed shows higher accuracy for both cases. }

{\change 
Figure~\ref{fig:latency} illustrates the communication latency of ChannelCompFed in comparison to analog and orthogonal communication methods for broadband communication with a bandwidth of $B = 1$ kHz. The division of the number of model parameters $N$ by $B$ results in communication rounds taking place across $\lceil N/B \rceil$ time slots. We employ the orthogonal frequency division multiple access (OFDMA) technique for the orthogonal communication, allocating a sub-channel to each edge device for model parameter transmission. As expected, both the analog scheme and ChannelCompFed achieve significantly ($10^5$ times) lower latency than the conventional orthogonal OFDMA approach. Note that latency is computed based on the rate-distortion function for a fading channel in~\cite{hekland2009shannon}, and the detailed analysis is provided in Appendix~\ref{Ap:Latency}. 
}

We observe that our proposed ChannelCompFed scheme enables existing modulation to train the model over the \ac{MAC} with similar learning performance for different numbers of antennas while it keeps the benefit of the analog scheme, such as spectral efficiency and low latency communication.

\input{fig/Fig_Latency}

\section{Conclusion}\label{sec:Conclusion}

In this study, we investigated \ac{FEEL} over a fading \ac{MAC}. To alleviate the communication burden, we considered a novel digital OAC scheme, termed ChannelComp, utilizing \ac{QAM} modulations. This method enables an ultra-low latency communication strategy for addressing the FEEL problem.

Our approach employs multiple antennas at the \ac{ES} to counteract the detrimental impacts of fading effects inherent in wireless channels. Our theoretical analysis rigorously determined the number of antennas needed at the ES to mitigate the fading effects effectively, shedding light on systems design considerations.

We extended our study to derive our proposed scheme's \ac{MSE} under noisy and fading MAC conditions. Capitalizing on the obtained MSE expressions, we proved the convergence rate for a non-convex loss function, offering insights into the algorithm's behavior under noisy and fading scenarios.

Our numerical results corroborate the theoretical conclusions concerning \ac{MSE} and the convergence properties of the digital \ac{FEEL} framework. These results showed that an enhanced estimation of average local model updates is achievable by augmenting the antenna number at the \ac{ES} and employing higher-order \ac{QAM}, improving the model accuracy up to $60\%$.

In conclusion, this work introduces a viable approach to implementing \ac{FEEL} over fading \ac{MAC} and provides rigorous analytical insights and empirical evidence to support its potential as a robust and efficient solution for distributed machine learning in wireless edge computing environments.

\appendix

\subsection{Proof of Proposition \ref{prop:EncDec}}\label{sec:ProofDeCode}

We first obtain the complex value of $\hat{g}:= \sum_{k=1}^K\mathscr{E}_q(\tilde{g}_k)/K$. Then,  we apply the decoder function $\mathscr{D}_{q}$ to the resultant value. In particular, for the real part of the average of encoded gradients, we have 
\begin{align}
\nonumber
  \mathcal{R}(\mathfrak{Re}(\hat{r})) & = \frac{1}{K}\mathcal{R}\Big(\sum\nolimits_{k=1}^K [\mathscr{E}_{q}(\tilde{g}_{k})]_1\Big),\\
   & \nonumber = \frac{1}{K}\mathcal{R}\bigg(\sum\nolimits_{k=1}^K \tilde{g}_{k} -2^{b} \sum\nolimits_{k=1}^K \left\lfloor\frac{\tilde{g}_{k}}{2^{b}}\right\rfloor\bigg) + \frac{2^{b}-1}{2}, \\ \label{eq:encode1}
    & =\mathcal{R}(\tilde{g}) +  \frac{2^{b}-1}{2} -  \frac{2^{b}}{K}\sum\nolimits_{k=1}^K \left\lfloor\frac{\tilde{g}_{k}}{2^{b}}\right\rfloor. 
\end{align}
For the imaginary part of $\hat{r}$,
\begin{align}
    \label{eq:encode2}
    \mathfrak{Im}(\hat{r}) =  \frac{1}{K}\mathcal{R}\Big(\sum_{k=1}^K [\mathscr{E}_{q}(\tilde{g}_{k})]_2\Big) & =   \frac{1}{K}\sum_{k=1}^K \left\lfloor \tilde{g}_{k}2^{-b}\right\rfloor +\frac{2^{b}-1}{2}.
\end{align}
Then, by substituting Eqs.~\eqref{eq:encode1} and \eqref{eq:encode2} into Eq.~\eqref{eq:decode}, we have
\begin{align}
    \label{eq:gnDE}
    \frac{1}{K}\mathscr{D}_{q}\Big(\sum\nolimits_{k=1}^K \mathscr{E}_{q}(\tilde{g}_{k})\Big) = \tilde{g}, 
\end{align}
This concludes the proof.

\subsection{Proof of Proposition~\ref{Pr:MSE}}\label{sec:ProofMSE}

Recall that $g^n$, $\tilde{g}^n$ and $\hat{g}^n$ are the $n$-th element of gradient $\bm{g}$, quantized gradient $\tilde{\bm{g}}$, and the estimated gradient $\hat{\bm{g}}$, respectively. We first split the error of the quantization and the channel using the triangular inequality on every element of the gradient as follows
\begin{align}
    |g^n- \hat{g}^n|^2 \leq |g^n- \tilde{g}^n|^2 + |\tilde{g}^n- \hat{g}^n|^2.
\end{align}
 We obtain upper bounds for each error term separately in the sequel. The expected value of the first term equals the variance of the quantization error, which is given by
\begin{align}
    \label{eq:UpperGradQuan}
    \mathbb{E}[|g^n- \tilde{g}^n|^2] \leq  \frac{1}{12K}\Big(\frac{2\Delta_g}{q}\Big)^2 = \frac{\Delta_g^2}{3Kq^2}. 
\end{align}
 Here, we obtained the variance under the assumption of a uniform distribution of the gradient elements. Then, for the second term, we follow the same procedure as the one used in \cite[Theorem 1]{razavikia2023SumCode}. For the received signal at \ac{ES} in Eq.~\eqref{eq:Aggnoisycase}, we can write 
\begin{align}
    \nonumber
    \hat{g}^n & = \frac{1}{K} \mathscr{D}_{q}\Big(\sum\nolimits_{k=1}^K \mathscr{E}_{q}(\tilde{g}_{k}^n) + \tilde{z}^n \Big),
     \\ &  = \tilde{g}^n + \frac{1}{K}\Big(\mathfrak{Re}(\tilde{z}^n) + 2^b\mathfrak{Im}(\tilde{z}^n)\Big).
\end{align}
Then, we can write the \ac{MSE} as follows
\begin{align}
    \nonumber
    \mathbb{E}[|\tilde{g}^n-\hat{g}^n|_2^2] & = \frac{1}{K^2}\mathbb{E}[|\mathfrak{Re}(\tilde{z}^n)|_2^2] + \frac{2^{2b}}{K^2}\mathbb{E}[|\mathfrak{Im}(\tilde{z}^n)|_2^2],
\end{align}
where the equality is due to that $\mathfrak{Re}(\tilde{z}^n)$ and $\mathfrak{Im}(\tilde{z}^n)$ are independent random variables. Furthermore, since the noise $\tilde{z}^n$ is a circular AWGN, the variance of the noise is the same for both real and imaginary parts. Thereafter, we define $e_r := |\mathfrak{Re}(\tilde{z}^n)|_2^2 = |\mathfrak{Im}(\tilde{z}^n)|_2^2$, which leads to
\begin{align}
    \label{eq:Egghat}
    \mathbb{E}[|\tilde{g}^n-\hat{g}^n|_2^2] & = \frac{(1+q)}{K^2}\mathbb{E}[e_r].
\end{align}
Hence, we only require computing the term $e_r$. The noise $\tilde{z}^n$ has Normal distribution  $\mathcal{CN}(0,\sigma_z^2/\sqrt{N_r})$, so we can compute the expected value for $e_r$ conditioned on $\mathfrak{Re}(\tilde{z}^n)$ or $\mathfrak{Im}(\tilde{z}^n)$ being between certain decision boundaries. Let $P_{\ell}$ be the probability that error becomes $\ell$, i.e., $P_{\ell}: = \Pr(e_r = \ell)$, then we have from \cite[Section 6.1.4]{goldsmith2005wireless},
\begin{align}
\nonumber
    P_{\ell} = 2\Big(1 - \frac{\ell}{2^b}\Big)\Big(Q\Big(\frac{(2\ell-1)\sqrt{N_r}}{\sigma_z}\Big) - Q\Big(\frac{(2\ell+1)\sqrt{N_r}}{\sigma_z}\Big)\Big).
\end{align}
Then, the expected value of the error is computed as
\begin{align*}
     \nonumber \mathbb{E}[e_{r}] &=  \sum\nolimits_{\ell=1}^{2^b-1}\ell^2\times P_{\ell}, \\ & = 2 \sum\nolimits_{\ell=1}^{2^b-1}\Big(2\ell-1 \frac{3\ell-3\ell^2 -1}{2^b}\Big) Q\Big(\frac{(2\ell-1)\sqrt{N_r}}{\sigma_z}\Big).
\end{align*}
Next, we use the expression in Eqs.~\eqref{eq:Egghat} and \eqref{eq:UpperGradQuan} to obtain an upper bound on the gradient average as follows.
\begin{align}
    \mathbb{E}[\|\bm{g}-\hat{\bm{g}}\|^2] \leq \frac{N(1+q)e_r}{K^2} + \frac{N\Delta_g^2}{3Kq^2}.
\end{align}
Finally, we conclude the proof.

\subsection{Proof of Theorem \ref{th:NRerror}}\label{sec:proofNRth}
We define the error as $e^n:= s^{n}-\hat{s}^n$, then use Eq.~\eqref{eq:sigrec} to expand the error as follows
\begin{align}
\label{eq:errorn}
    {e}^{n}  =  e_{\rm sig}^{n} + e_{\rm int}^{n} + e_{\rm noise}^{n},
\end{align}
where the terms are defined as
\begin{subequations}
    \label{eq:Signalvalues}
    \begin{align}
    \label{eq:sig_error}
    e_{\rm sig}^{n} &:=  \sum\nolimits_{k=1}^{K}\Big(\frac{\|\bm{h}_{k}^n\|^2}{\sigma_h^2N_r}-1\Big){s}_{k}^n,  \\ \label{eq:Interfence}
     e_{\rm int}^{n}&:= \sum\nolimits_{k,k',k\neq k'}^{K}\frac{\langle \bm{h}_{k}^n, \bm{h}_{k'}^n\rangle}{N_r \sigma_h^2} {s}_{k'}^{n}, \\  \label{eq:Noiseterm}
     e_{\rm noise}^{n} &:=  \sum\nolimits_{k=1}^{K}\frac{\langle \bm{h}_{k}^n, \bm{z}_{n}\rangle}{\sigma_h^2N_r}.
\end{align}
\end{subequations}
Hence, the terms  $e_{\rm sig}^{n},e_{\rm int}^{n}$ and  $e_{\rm noise}^{n}$ are random variables with the following first and second order moments: 
\begin{subequations}
    \begin{align}
    & \mathbb{E}[e_{\rm sig}^{n}] = 0, \quad  \mathbb{E}[|e_{\rm sig}^{n}|^2] = \alpha_1^n/N_r, \\
    & \mathbb{E}[e_{\rm int}^{n}] = 0, \quad  \mathbb{E}[|e_{\rm int}^{n}|^2]  = \frac{(K-1)\alpha_2^n}{N_r},\\
    & \mathbb{E}[e_{\rm noise}^{n}] = 0,\quad  \mathbb{E}[|e_{\rm noise}^{n}|^2] = K \frac{\sigma_z^2}{\sigma_h^2N_r},
\end{align}
\end{subequations}
where $\alpha_1^n:= 2\sum_{k=1}^K|s_k^n|^2, \alpha_2^n:= \sum_{k,k',k\neq k'}^{K} |s_k^ns_{k'}^n|,$ and the expectations are over noise and channel coefficients distributions. In the following, we show how fast the tails of these random variables deviate from their expected values. To this end, we bring some relevant definitions and properties with Bernstein’s inequality for obtaining upper bounds on their concentration. 
\begin{definition}
    For a random variable $X$ with $\mathbb{E}[X] = 0$, we define 
    \begin{align}
        \lambda{(X)}:= \inf \{ & t > 0: \mathbb{E}[{\rm exp}(|X|/ t)\leq 2\}. 
    \end{align}
    $X$ is called a sub-exponential random variable with parameter $\lambda$,  if $\lambda(X)<\infty$.  
\end{definition}
\begin{lem}(\hspace{-0.1pt}\cite[Theorem 2.8.1]{vershynin2020high})\label{lem:Bern}
    Let $X_1, \ldots, X_n$ be independent random variables such that $\mathbb{E}[X_i]=0$  and sub-exponential random variables with parameter $\lambda$. Then for any $t > 0$, we have
    \begin{align}
         \mathbb{P}\bigg[\Big|\frac{1}{n}\sum\nolimits_{i=1}^nX_i\Big|>t\bigg] \leq 2{\rm exp}\Big(-\frac{n}{2}\min\Big(\frac{t^2}{\lambda^2},\frac{t}{\lambda}\Big) \Big).
    \end{align}
\end{lem}
 In the sequel, we show that each error term in Eq.~\eqref{eq:Signalvalues}, vanishes for a large enough number of antennas $N_r$ using Lemma \ref{lem:Bern}. For the first error term, $e_{\rm sig}^n$, we have the following Lemma. 
\begin{lem}\label{lem:Delta1}
    Let $\epsilon_1>0$ and $\delta_1>0$ be positive scalars. Then,  the absolute value of the signal error in Eq.~\eqref{eq:sig_error}, $|e_{\rm sig}^n|$, is upper bounded by the scalar $\epsilon_1$ with probability at least $1-\delta_1$, if 
    \vspace{-5pt}
\begin{align}
   N_r  \geq  \frac{4\gamma_n^2}{\epsilon_1^2}\ln{\Big(\frac{2K}{\delta_1}\Big)},
\end{align}
    where $\gamma_n = \sum_{k}^{K}|{s}_{k}^{n}|$.
\end{lem}
\begin{proof}
    See Appendix \ref{sec:lemDelta1}.
\end{proof}

\begin{lem}\label{lem:Delta2}
    Let $\epsilon_2>0$ and $\delta_2>0$ be positive scalars. Then,  the absolute value of the interference error term in Eq.~\eqref{eq:Interfence}, $|e_{\rm int}^{n}|$, is upper bounded by the scalar $\epsilon_2$ with probability at least $1-\delta_2$, if 
    \vspace{-5pt}
\begin{align}
    N_r \geq \frac{8(K-1)^2\gamma^2_n}{\epsilon^2_2}\ln{\Big(\frac{2(K-1)}{\delta_2}\Big)},
\end{align}
    where $\gamma_n = \sum_{k}^{K}|{s}_{k}^{n}|$.
\end{lem}
\begin{proof}
    See Appendix \ref{sec:lemDelta2}.
\end{proof}

\begin{lem}\label{lem:Delta3}
    Let $\epsilon_3>0$ and $\delta_3>0$ be positive scalars. Then, the absolute value of the noise error in Eq.~\eqref{eq:Noiseterm}, $|e_{\rm noise}^n|$, is upper  bounded by the scalar $\epsilon_3$ with probability at least $1-\delta_3$, if 
\begin{align}
    N_r \geq \frac{8\sigma_z^2K^2}{\sigma_h^2\epsilon_3^2}\ln{\Big(\frac{2(K-1)}{\delta_3}\Big)},
\end{align}
    where $\gamma_n = \sum_{k}^{K}|{s}_{k}^{n}|$.
\end{lem}
\begin{proof}
    See Appendix \ref{sec:lemDelta3}.
\end{proof}
From Lemmas \ref{lem:Delta1}, \ref{lem:Delta2}, and \ref{lem:Delta3}, 
we can bound the  absolute value of the defined error in Eq.~\eqref{eq:errorn}, as follows
\begin{align}
 \nonumber
    |e^{n}| & = |e_{\rm sig}^n + e_{\rm int}^n + e_{\rm noise}^n| \leq |e_{\rm sig}^n| + |e_{\rm int}^n| + |e_{\rm noise}^n|, \\
     & \leq \epsilon_1 + \epsilon_2 + \epsilon_3 \leq \epsilon.
\end{align}
Due to the union bound, the events of Lemmas \ref{lem:Delta1}, \ref{lem:Delta2}, and \ref{lem:Delta3} hold simultaneously with probability no less than $1-(\delta_1 + \delta_2 +\delta_3)$ or $1-\delta$, where $\delta \geq \delta_1 + \delta_2 +\delta_3$~\cite{Razavikia2019Hankel,bokaei2023harmonic}. Thus, this proves Eq.~\eqref{eq:Expecepsilonupp}. 

To satisfy all the constraints in Lemmas \ref{lem:Delta1}, \ref{lem:Delta2}, and \ref{lem:Delta3}, the number of antennas needs to be greater than all the inequalities, i.e., 
\begin{align}
    \nonumber
    N_r \geq  & \max\Big\{\frac{4\gamma_n^2}{\epsilon_1^2}\ln{\Big(\frac{2K}{\delta_1}\Big)}, \frac{8(K-1)^2\gamma^2_n}{\epsilon^2_2}\ln{\Big(\frac{2(K-1)}{\delta_2}\Big)},\\ & \frac{8\sigma_z^2K^2}{\sigma_h^2\epsilon_3^2}\ln{\Big(\frac{2(K-1)}{\delta_2}\Big)} \Big\}.
\end{align}
For further simplifications, we can set $\delta_1 = \delta/3$, $\delta_2 = {(1-1}/{K})\delta/3$, and $\delta_3= {(1-1}/{K})\delta/3$, which leads to 
\begin{align}
 \delta_1 + \delta_2 +\delta_3 \leq  \frac{\delta}{3} (1 + 1 + 1 - \frac{2}{K}) \leq \delta.
\end{align}
Moreover, we obtain the following lower bound for $N_r$: 
\begin{align}
    \nonumber
    N_r \geq  \max\Big\{\frac{4\gamma_n^2}{\epsilon_1^2}, \frac{8(K-1)^2\gamma^2_n}{\epsilon^2_2}, \frac{8\sigma_z^2K^2}{\sigma_h^2\epsilon_3^2} \Big\} \ln{\Big(\frac{6K}{\delta}\Big)}.
\end{align}
Next, let us rewrite  $\epsilon_1$, $\epsilon_2$, and $\epsilon_3$ as follows.
\begin{subequations}
    \begin{align}
   \epsilon_1 & = \frac{\sigma_h}{\sigma_zK}\epsilon'', \\
   \epsilon_2 & = \frac{\sigma_h(K-1)}{\sqrt{2}\sigma_zK}\epsilon'', \\
   \epsilon_3 & = \frac{1}{\sqrt{2}\gamma_n}\epsilon'', 
\end{align}
\end{subequations}
for some positive value $\epsilon''>0$. Hence, we have that $ \epsilon_1 + \epsilon_2 + \epsilon_3 <  \epsilon''(\frac{\sigma_h}{\sqrt{2}\sigma_z}+ \frac{1}{\sqrt{2}\gamma_n})$. Then, by setting $\epsilon = \epsilon''c_n/\sqrt{2} $, where $c_n = {\sigma_h}/{\sigma_z}+ {1}/{\gamma_n}$,  we obtain the following expression for $N_r$:
\begin{align}
    \nonumber
    N_r \geq  \frac{8\gamma_n^2\sigma_z^2K^2}{\sigma_h^2\epsilon^2c_n^2} \ln{\Big(\frac{6K}{\delta}\Big)}.
\end{align}
This proves Eq.~\eqref{eq:UpperboundNR}. Then, for the expected value in Eq.~\eqref{eq:Expecepsilonupp}, from \cite{tropp2015introduction}[Theorem 1.6.2], we know that the expected values of the bounded random variable, $e^n$, can also be bounded as follows:
\begin{align}
  \mathbb{E}[|e^n|]  \leq \frac{4K\gamma_n}{\sqrt{N_r}c_n} (\sqrt{\pi} +\ln{(6K)}). 
\end{align}
Thus, we conclude the proof for Theorem~\ref{th:NRerror}.

\subsection{Proof of Lemma \ref{lem:Delta1}}\label{sec:lemDelta1}

We first define the variable $Y_\ell$ for $\ell \in [N_r]$ as follows:
\begin{align}
     Y_\ell^{k}  :=  \frac{1}{\sigma^2_h}(h_{k,\ell}^{n})^2, \quad {\forall} k \in [K], \quad n \in [N],
\end{align}
which yields $\sum_{\ell=1}^{N_r}Y_\ell^{k}/N_r = \|\bm{h}_k^{n}\|^2/{N_r\sigma_h^2}$. Due to the Normal distribution assumption on $h_{k,\ell}^{n}$ from Section~\ref{sec:Commodel}, $Y_{\ell}^k$  is  sub-exponential with
parameters $(2,4)$.   Consequently, since the variables $Y_{\ell}^{k}$ are independent for $(\ell, k) \in [N_r]\times[K]$, the average  $\sum_{\ell=1}^{N_r}Y_\ell^{k}/N_r$ has Chi-squared distribution with $N_r$ degrees of freedom. To bound this variable, we use Lemma \ref{lem:Laurent}.

\begin{lem}\cite[Lemma 1]{Laurent2000estimation}\label{lem:Laurent}
    Let $(Y_1,\ldots, Y_n)$ be \ac{iid} Normal random variables. We set $Y = \sum_{i=1}^{n}Y_i^2-n$. Then, the following inequalities hold for any positive $t$:
    \begin{subequations}
           \begin{align}
        \mathbb{P}\Big(Y \geq 2\sqrt{nt} + 2t \Big) & \leq \exp{(-t)},\\
        \mathbb{P}\Big(Y \leq - 2\sqrt{nt} \Big) & \leq \exp{(-t)}.
        \end{align}
    \end{subequations}
\end{lem}
By applying Lemma \ref{lem:Laurent}, the variable $\sum_{\ell=1}^{N_r}Y_\ell^k/N_r$ is bounded below as 
\begin{align}
    \label{eq:Y1lower}
    \bigg|\frac{\sum_{\ell=1}^{N_r}Y_\ell^k}{N_r} - 1 \bigg| \geq c_1,\quad k\in [K],
\end{align}
with probability at most $ \exp{(-N_rc_1^2/4)} + \exp{(-N_rc_1^2/2)} \leq 2\exp{(-N_rc_1^2/4)}$ for each term. Invoking the union bound, the probability that at least one of the events in  Eq.~\eqref{eq:Y1lower} occurs is at most  
\begin{align}
\nonumber
\sum_{k=1}^K\Pr\bigg(\bigg|\frac{\sum_{\ell=1}^{N_r}Y_\ell^k}{N_r} - 1 \bigg| \geq c_1\bigg) \leq  K \times 2\exp{(-N_rc_1^2/4)}.
\end{align}
Hence, all the terms for the variable $\sum_{\ell=1}^{N_r}Y_\ell^k/N_r$ are bounded above as 
\begin{align}
    \label{eq:Y1upper}
    \bigg|\frac{\sum_{\ell=1}^{N_r}Y_\ell^k}{N_r} - 1 \bigg| \leq c_1,\quad k\in [K],
\end{align}
with probability at least $1- 2K\exp{(-N_rc_1^2/4)}$. 
Accordingly, for the error $e_{\rm sig}^n$, we can use Eqs.~\eqref{eq:Signalvalues} as follows:
\begin{align}
\nonumber
   |s^{n} - s_{\rm sig}^{n}| & =  \bigg|\sum_{k=1}^{K}\Big(\frac{\sum_{\ell=1}^{N_r}Y_\ell^k}{N_r} - 1\Big)s_k^{n} \bigg|, \\ \nonumber
    & \leq  \bigg|\sum_{k=1}^{K}\Big|\frac{\sum_{\ell=1}^{N_r}Y_\ell^k}{N_r} - 1\Big|s_k^{n} \bigg| \leq c_1 \sum_{k=1}^{K}|s_k^{n}|, \\
    & = c_1 \gamma_n,
\end{align}
where the first inequality is the triangle inequality, and the second inequality uses the upper bound in Eq.~\eqref{eq:Y1upper}. Also, $\gamma_n$ is defined as
\begin{align}
    \label{eq:Gammadef}
    \gamma_n : = \sum\nolimits_{k=1}^{K}|s_k^{n}|.
\end{align}
To bound  $e_{\rm sig}^n$ from above by $\epsilon_1$, we have to bound $c_1$ to be  $ c_1\gamma_n\leq \epsilon_1$, or, equivalently, $c_1 \leq \epsilon_1/\gamma_n$.    
As a result, the error  term is upper bounded by $\epsilon_1$, i.e., $e_{\rm sig}^n \leq \epsilon_1$,  with probability at least $1-\delta_1$ as long as the following inequality on $\delta_1$ is satisfied:
\begin{align}
\label{eq:c_1N_r}
    K \times 2\exp{(-N_rc_1^2/4)} & \leq \delta_1. 
\end{align}
Hence, we rearrange the expression in Eq.~\eqref{eq:c_1N_r} to have $N_r$ on the left side hand of the inequality, i.e., 
\begin{align}
    N_r & \geq  \frac{4\gamma_n^2}{\epsilon_1^2}\ln{\Big(\frac{2K}{\delta_1}\Big)},
\end{align}
where the last inequality is fulfilled due to $c_1$  being upper bounded by $\epsilon_1/\gamma_n$. Therefore, the proof is concluded.

\subsection{Proof of Lemma~\ref{lem:Delta2}}\label{sec:lemDelta2}

Let us define the variable $X_\ell$ for $\ell \in [N_r]$ as follows,
\begin{align}
    \label{eq:defXlhkhk'}
     X_\ell^{(k,k')}  :=  \frac{1}{\sigma^2_h}h_{k,\ell}^nh_{k',\ell}^n, \quad {\text{for}}~~k,k' \in [K], \quad n \in [N].
\end{align}
Consequently, we have 
\begin{align}
    \frac{1}{N_r}\sum\nolimits_{\ell=1}^{N_r}X_\ell^{(k,k')} = \frac{1}{N_r\sigma^2_h} \langle \bm{h}_{k}^n, \bm{h}_{k'}^n\rangle.
\end{align}
Considering that $h_{k,\ell}^n/\sigma_h$ and $h_{k',\ell}^n/\sigma_h$ are Normal random variables, the product of them in Eq.~\eqref{eq:defXlhkhk'} makes $X_\ell^{(k,k')}$ sub-exponential with $\lambda = 2$~\cite[Lemma 2.7.7]{vershynin2020high}.  Therefore,  the direct results of applying Lemma \ref{lem:Bern} to the term $\frac{1}{N_r\sigma^2_h} \langle \bm{h}_{k}^n, \bm{h}_{k'}^n\rangle$, gives us the following upper bound
\begin{align}
    \label{eq:BoundX}
    \Big| \frac{1}{N_r}\sum\nolimits_{\ell=1}^{N_r}X_\ell^{(k,k')} \Big| \leq c_2,
\end{align}
with the probability at least $1- 2{\rm exp}(-N_rc_2^2/8)$. Hence, to make sure that $|s^{n}_{\rm int}|$ is bounded by $\epsilon_2$ with probability  at least $1-\delta_2$, we have 
\begin{align}
\nonumber
     |s^{n}_{\rm int}| & = \Big|\sum_{k,k',k\neq k'}^{K}\frac{\langle \bm{h}_{k}^n, \bm{h}_{k'}^n\rangle}{N_r \sigma_h^2} {s}_{k'}^{n}\Big| \leq \sum_{k,k',k\neq k'}^{K}\Big|\frac{1}{N_r}X_{\ell}^{(k,k')}\Big||{s}_{k'}^{n}|, \\
      & \leq (K-1)c_2\sum\nolimits_{k}^{K}|{s}_{k}^{n}|.
\end{align}
Then, let us define $\gamma_n : = \sum_{k}^{K}|{s}_{k}^{n}|$, such that
\begin{align}
    \label{eq:epsilonp}
    (K-1)c_2\gamma_n  \leq  \epsilon_2 \Rightarrow  c_2 \leq \frac{\epsilon_2}{(K-1)\gamma_n}.
\end{align}
Using the same arguments from Appendix~\ref{sec:lemDelta1},  the condition  in Eq.~\eqref{eq:BoundX} can be satisfied for all $k,k'\in [K]$ with probability at least 
\begin{align}
   1 - (K-1)\times2 \exp{(-N_rc_2^2/8)}.
\end{align}
Then, the number of required antennas $N_r$ needs to fulfill the inequality below:
\begin{align}
\nonumber
    (K-1)\times2 \exp{(-N_rc_2^2/8)} & \leq \delta_2, \\ \nonumber
    -N_rc_2^2/8 & \leq \ln{\Big(\frac{\delta_2}{2(K-1)}\Big)}, \\ \label{eq:NrLnone}
    N_r & \geq \frac{8}{c_2^2}\ln{\Big(\frac{2(K-1)}{\delta_2}\Big)}.
\end{align}
Afterward, by substituting  Eq.~\eqref{eq:epsilonp} into Eq.~\eqref{eq:NrLnone}, we obtain
\begin{align}
\label{eq:Nrdelta2}
 N_r \geq \frac{8(K-1)^2\gamma^2_n}{\epsilon^2_2}\ln{\Big(\frac{2(K-1)}{\delta_2}\Big)}.
\end{align}
Using Eq.~\eqref{eq:Nrdelta2}, we conclude the proof.

\subsection{Proof of Lemma \ref{lem:Delta3}}\label{sec:lemDelta3}

Let $W_{\ell}$ be defined as 
\begin{align}
    W_{\ell} := \frac{h_{k,\ell}^n}{\sigma_h} \frac{z_{\ell}^{n}}{\sigma_h}, \quad \forall k \in [K], n\in [N],
\end{align}
where $\ell \in [N_r]$. Here, $h_{k,\ell}^n/\sigma_h$ and $ {z_{\ell}^{n}}/{\sigma_h}$ are two sub-Gaussian random variables with parameters $1$ and $ \sigma_z^2/\sigma_h^2$, respectively. Thus, $W_{\ell}$ is a sub-exponential random variable with parameter $\lambda = 2\times 1 \times \sigma_z / \sigma_h$ \cite{wainwright2019high}. Similar to Eq.~\eqref{eq:BoundX} in Appendix~\ref{sec:lemDelta2}, we apply Lemma \ref{lem:Bern} to bound the summation  $\sum_{\ell}^{N_r}W_{\ell} $ as follows
\begin{align}
    \Big| \frac{1}{N_r}\sum\nolimits_{\ell=1}^{N_r}W_{\ell}\Big| =\Big| \frac{1}{N_r\sigma_h^2}\langle \bm{h}_{k}^n, \bm{z}_{n}\rangle\Big|  \leq c_3,
\end{align}
with probability no less than $ 1 - 2\exp{(-N_r\sigma_h^2c_3^2/8\sigma_z^2)}$. Therefore, $e_{\rm noise}^{n}$ is bounded from above as follows
\begin{align}
    |e_{\rm noise}^{n}|  =  \Big|\sum\nolimits_{k=1}^{K}\frac{\langle \bm{h}_{k}^n, \bm{z}_{n}\rangle}{\sigma_h^2N_r}\Big| \leq K c_3 \leq \epsilon_3.
\end{align}
Hence, $c_3 \leq \epsilon_3/K$.

Then, by using the union bound similar to Appendix~\ref{sec:lemDelta1}, for all $k$ items of $e_{\rm noise}^{n}$, we have $2K \exp{(-N_r\sigma_h^2c_3^2/8\sigma_z^2)}$ to be lower than $\delta_3$, which leads to 
\begin{align}
   N_r & \geq \frac{8\sigma_z^2}{\sigma_h^2c_3^2}\ln{\Big(\frac{2(K-1)}{\delta_3}\Big)}.  
\end{align}
By setting $c_3 = \epsilon_3/K$, we obtain  
\begin{align}
   N_r & \geq \frac{8\sigma_z^2K^2}{\sigma_h^2\epsilon_3^2}\ln{\Big(\frac{2(K-1)}{\delta_3}\Big)}.  
\end{align}
Thus, we can conclude the proof.

\subsection{Proof of Proposition~\ref{cor:norm2}}\label{sec:APMSE}

To obtain an upper bound on the \ac{MSE} of $\bm{g}$, we use the triangular inequality to split the channel and quantization errors as follows: 
\begin{align}
    \label{eq:gtwohatandtild}
    \| \bm{g} - \hat{\bm{g}}\|^2 \leq  \| \bm{g} -  \tilde{\bm{g}} \|^2  + \| \tilde{\bm{g}} - \hat{\bm{g}}\|^2, 
\end{align}
where the first error term is due to the quantization error, and the second term represents the channel errors. Considering  the fact that the elements of gradient $\bm{g}$ are bounded by value $\Delta_q$, the expected value of the first term is bounded as
\begin{align}
    \mathbb{E}[\| \bm{g} -  \tilde{\bm{g}} \|^2] \leq \frac{N\Delta_q^2}{3Kq^2}. 
\end{align}
To upper bound the absolute error of the gradient $\tilde{\bm{g}}$ and $\hat{\bm{g}}$, we first need to obtain an upper bound on the error of every element. Due to the channel fading and noise, the transmitted gradient over sub-channel $n$-th, i.e.,  $\tilde{g}_{k}^{n}$ in Eq.~\eqref{eq:gnDE} is contaminated with error. Hence, the estimated gradient can be written as
\begin{align}
    \nonumber
    \hat{g}^n & =  \frac{1}{K}\mathscr{D}_{q}\Big(\sum\nolimits_{k=1}^K \mathscr{E}_{q}(\tilde{g}_{k}^n) + e_{ch} \Big),\\ \nonumber
    & =  \frac{1}{K}\mathfrak{Re}{\Big(\sum\nolimits_{k=1}^K \mathscr{E}_{q}(\tilde{g}_{k}^n) + e_{ch}\Big)} \\ &  + \frac{2^{b}}{K} \left(\mathfrak{Im}{\Big(\frac{1}{K}\sum_{k=1}^K \mathscr{E}_{q}(\tilde{g}_{k}^n) + e_{ch}\Big)} + \tfrac{2^{b}+1}{2}\right) + \tfrac{2^{b}-1}{2}, \nonumber \\
    & = \tilde{g}^n + \frac{1}{K}\mathfrak{Re}{(e_{ch})} + \frac{2^{b}}{K} \cdot \mathfrak{Im}{(e_{ch})}.  \label{eq:gREIM2}
\end{align}
As a result, for the error $e_g : = | \hat{g}^n -  \tilde{g}^n|$, we have
\begin{align}
    \nonumber
    | \hat{g}^n -  \tilde{g}^n| & = \frac{1}{K}|\mathfrak{Re}{(e_{ch})} + 2^{b} \cdot\mathfrak{Im}{(e_{ch})}|, \\
    & \leq  \frac{1}{K}|\mathfrak{Re}{(e_{ch})} |  + \frac{2^{b}}{K}|\mathfrak{Im}{(e_{ch})}|.
\end{align}
Then, for $N_r \geq \frac{8\sigma_z^2K^2}{\sigma_h^2\epsilon^2}\ln{\big(\frac{6K}{\delta}\big)}$ Theorem~\ref{th:NRerror}, we can bound the error as follows
\begin{align}
    \label{eq:gghaterror}
    | \hat{g}^n -  \tilde{g}^n| \leq \frac{1}{K}(\epsilon + 2^{b}\epsilon) = \frac{(2^{b}+1)}{K}\epsilon.
\end{align}
Afterward, we use Eq.~\eqref{eq:gghaterror} to obtain an upper bound on the whole gradient as 
\begin{align}
    \label{eq:ghatgtild}
    \|\tilde{\bm{g}} - \hat{\bm{g}}\|^2 & \leq N(2^{2b}+1 +2^{b+1} )\frac{\epsilon^2}{K^2} \leq 2Nq\frac{\epsilon^2}{K^2}. 
\end{align}
By defining $\epsilon_{\rm fad}^2 := 2Nq \epsilon^2/K^2$, we can complete the proof for Eq.~\eqref{eq:ErrorProb}. Next, to obtain an upper bound for the expected value, we use the same argument in Eq.~\eqref{eq:gghaterror} from Appendix~\ref{sec:proofNRth} for bounded variables in $| \hat{g}^n -  \tilde{g}^n|$, which leads to the following: 
\begin{align}
  \mathbb{E}[|\hat{g}^n-g^n|]  \leq 4\frac{(\sqrt{q}+1)\gamma_n}{\sqrt{N_r}c_n} (\sqrt{\pi} +\ln{(6K)}). 
\end{align}
Hence, for the norm of $\bm{g}-\hat{\bm{g}}$, we have 
\begin{align}
   \nonumber
  \mathbb{E}[\|\bm{g}-\hat{\bm{g}}\|^2]  & \leq 16\frac{q N \max_{n}\gamma_n^2}{N_r\min_n{c_n^2}} (\pi +2\ln{(6K)}^2), \\
  & \leq 16\frac{qN\gamma_{\rm max}^2 }{N_rc_{\rm min}^2} (\pi +2\ln{(6K)}^2). \label{eq:upperonghat}
\end{align}
Finally, by taking expectation over both sides of Eq.~\eqref{eq:gtwohatandtild} and substituting the upper bound in Eq.~\eqref{eq:upperonghat}, we obtain 
\begin{align}
    \mathbb{E} [\| \bm{g} - \hat{\bm{g}}\|^2] \leq \sigma_q^2 + 16\frac{qN\gamma_{\rm max}^2 }{N_rc_{\rm min}^2} (\pi +2\ln{(6K)}^2). 
\end{align}
This concludes the proof.

\subsection{Proof of Proposition \ref{prop:ConvAWGN}}\label{Ap:ConvAWGN}

The proof herein follows the same procedure used in \cite{csahin2023over,bernstein2018signsgd,zhu2020one}. By using Assumption~\ref{as:smooth}, we can write the following upper bound for the loss functions of two consecutive communication rounds: 
\begin{align}
    \nonumber
    \mathcal{L}(\bm{g}{(m+1)})-\mathcal{L}(\bm{g}{(m)}) \leq\hspace{-2pt} -\eta \langle \bm{g}{(m)}, \hat{\bm{g}}(m)\rangle \hspace{-2pt}+ \hspace{-2pt}\frac{\eta^2L}{2}\|\hat{\bm{g}}{(m)}\|^2. 
\end{align}
Here, we have $\bm{w}{(m+1)} = \bm{w}{(m)} - \eta \hat{\bm{g}}{(m)}$. Taking the expectation from both sides of an inequality, we get the following 
\begin{align}
    \nonumber
    \mathbb{E}\big[\mathcal{L}(\bm{g}{(m+1)})-& \mathcal{L}(\bm{g}{(m)})\big] \leq -\eta \langle \bm{g}{(m)}, \mathbb{E}[\hat{\bm{g}}]\rangle \\ &\nonumber  + \frac{\eta^2L}{2} \mathbb{E}\big[\|\hat{\bm{g}}{(m)}\|^2\big], \\ \nonumber
    & = -\eta \|\bm{g}{(m)}\|^2 + \frac{\eta^2L}{2} \mathbb{E}\big[\|\hat{\bm{g}}{(m)}\|^2\big], \\ \nonumber
    & \leq -\eta \|\bm{g}{(m)}\|^2 + \frac{\eta^2L}{2} \big((\sigma_{\rm ch}^2+\sigma_{q}^2)+\bar{\theta}\big),
\end{align}
where the first equality is due to Assumption~\ref{as:UQ}, and the inequality comes from the upper bounds on the variance of the quantization error and gradient variance in Assumptions~\ref{as:GD} and \ref{as:UQ}, respectively. Also, to obtain the inequality $\mathbb{E}[\| \bm{g} -  \tilde{\bm{g}} \|^2] \leq \frac{N\Delta_q^2}{Kq^2}$, we used Popoviciu's inequality~\cite{bhatia2000better} for the variance of bounded variables, $|g^n_k|\leq \Delta_q$ for $k \in [K]$. Then, by performing a telescoping sum over the communication rounds and subtracting the $\mathcal{L}(\bm{g}^{*})$ from both sides of the inequality, we get the following. 
\begin{align}
    \nonumber
    \mathcal{L}(\bm{g}^{(1)})-\mathcal{L}(\bm{g}^{*})  \geq &  \mathcal{L}(\bm{g}{(1)})-\mathbb{E}[\mathcal{L}(\bm{g}{(T)})],  \\
      & \nonumber = \mathbb{E}\bigg[\sum_{m=1}^{T}\mathcal{L}(\bm{w}(m)) - \mathcal{L}(\bm{w}(m-1))\bigg], \\ 
      & \nonumber \geq \sum_{m=1}^{T} \mathbb{E}\bigg[\mathcal{L}(\bm{w}(m)) - \mathcal{L}(\bm{w}(m-1))\bigg]\\
      & \geq (-\eta + \frac{\eta^2L}{2})\mathbb{E}\Big[\sum\nolimits_{m=1}^{T}\|\bm{g}(m)\|^2\Big] \nonumber \\
      & +  \frac{\eta^2LT}{2}(\sigma_{\rm ch}^2 + \sigma_{q}^2 + \Bar{\theta}).
\end{align}
Finally, rearranging the last inequality concludes the proof.

\subsection{Latency Analysis}\label{Ap:Latency}

  {\change
 To derive the latency for the fading channel, the transmission rate-distortion function of the communication scheme is needed. In particular, to calculate the rate-distortion function for a fading channel with bandwidth $B$, taking into account an MSE distortion measure, the following formula from \cite{hekland2009shannon} is used:
\begin{align}
    \mathcal{R}(\sigma_d^2) = \max[B\ln\Big(\frac{\sigma_s^2}{\sigma_d^2}\Big),0],
\end{align}
where $\sigma_s^2$  is the signal power and $\sigma_d^2$ is the distortion. In our MIMO fading channel scenario in \eqref{eq:sigrec}, the distortion \(\sigma_d^2\) after beamforming and averaging is expressed as:
\begin{align}
    \sigma_{\rm d-Analog}^2&  = \frac{\sigma_z^2}{N_rK\sigma_h^2} + \frac{(K-1)\sigma_{\rm int}^2}{3N_r}, \\
    \sigma_{\rm s-Analog}^2 &= \sum_{k=1}^K|s_k^n|^2/K,
\end{align}
where $\sigma_{\rm int}^2= \sum_{k,k',k\neq k'}^{K} \mathbb{E}[|s_k^ns_{k'}^n|]$. Hence, to quantify the latency for analog OAC, the equation is formulated as follows:
\begin{align}
    T_{\text{ana}} := \frac{T_s N}{\mathcal{R}(\sigma_{\text{d-Analog}}^2)} = \frac{T_s}{B} \frac{N}{\ln( \sigma_{\text{s-Analog}}^2/\sigma_{\text{d-Analog}}^2)},
\end{align}
where \(T_s\) represents the duration of an OFDM symbol, and \(N\) indicates the model size, i.e., the number of model parameters.

For ChannelCompFed to adhere to a similar rate-distortion function, it experiences distinct distortions, described by:
\begin{subequations}
   \begin{align}
\label{eq:Sigm1}
    \sigma_{\text{d-CompFed}}^2 & = \frac{\Delta_q^2}{q^2K} + \frac{(1+q){\Tilde{e}_r}^2}{K^2}, \\
    \label{eq:Sigm2}
    \sigma_{\text{s-CompFed}}^2 &= \sum_{k=1}^K|s_k^n|^2/K \Big|\frac{2q^2}{q+1}\Big|,
\end{align} 
\end{subequations}
where \(\frac{\Delta_q^2}{q^2}\) represents the quantization error and \(\frac{2q^2}{q+1}\) arises from the fact that QAM modulation consumes \(\frac{2q^2}{q+1}\) less power than its analog counterpart. The error term $\Tilde{e}_r$ is the same as the error term defined in \eqref{eq:xerror} while the variance $\sigma_z$ needs to be replaced by $ \sigma_{\rm d-Analog}^2$. Consequently, the latency for ChannelCompFed is given by:
\begin{align*}
    T_{\text{CompFed}} := \frac{T_s N}{\mathcal{R}(\sigma_{\text{d-CompFed}}^2)} = \frac{T_s}{B} \frac{N}{\ln(\sigma_{\text{s-CompFed}}^2/\sigma_{\text{d-CompFed}}^2)},
\end{align*}
Thus, the latency reduction ratio of ChannelCompFed over its analog counterpart is defined as:
\begin{align}
    \gamma :=  \frac{T_{\text{CompFed}}}{T_{\text{ana}}} = \frac{\ln( \sigma_{\text{s-Analog}}^2/\sigma_{\text{d-Analog}}^2)}{\ln(\sigma_{\text{s-CompFed}}^2/\sigma_{\text{d-CompFed}}^2)}.
\end{align}
As \(q\) increases in \eqref{eq:Sigm1} and \eqref{eq:Sigm2}, the latency reduction ratio \(\gamma\) approaches one, indicating asymptotic performance equivalence. However, ChannelCompFed can facilitate low latency communication at low quantization levels, although with higher quantization errors.
}

\bibliographystyle{IEEEtran}
\bibliography{IEEEabrv,Ref}

\end{document}

%% file: acronyms.tex
\begin{acronym}[LTE-Advanced]
  \acro{2G}{second generation}
  \acro{ASK}{amplitude-shift keying}
  \acro{SumComp}{sum computation}
  \acro{6G}{sixth generation}
  \acro{3-DAP}{3-dimensional assignment problem}
  \acro{AA}{antenna array}
  \acro{AC}{admission control}
  \acro{AD}{attack-decay}
  \acro{ADC}{analog-to-digital conversion}
  \acro{ADMM}{alternating direction method of multipliers}
  \acro{ADSL}{asymmetric digital subscriber line}
  \acro{AHW}{alternate hop-and-wait}
  \acro{AI}{artificial intelligence}
  \acro{AirComp}{over-the-air computation}
  \acro{AM}{amplitude modulation}
  \acro{AMC}{adaptive modulation and coding}
  \acro{AP}{\LU{A}{a}ccess \LU{P}{p}oint}
  \acro{APA}{adaptive power allocation}
  \acro{ARMA}{autoregressive moving average}
  \acro{ARQ}{\LU{A}{a}utomatic \LU{R}{r}epeat \LU{R}{r}equest}
  \acro{ATES}{adaptive throughput-based efficiency-satisfaction trade-off}
  \acro{AWGN}{additive white Gaussian noise}
  \acro{BAA}{\LU{B}{b}roadband \LU{A}{a}nalog \LU{A}{a}ggregation}
  \acro{BB}{branch and bound}
  \acro{BCD}{block coordinate descent}
  \acro{BD}{block diagonalization}
  \acro{BER}{bit error rate}
  \acro{BF}{best fit}
  \acro{BFD}{bidirectional full duplex}
  \acro{BLER}{bLock error rate}
  \acro{BPC}{binary power control}
  \acro{BPSK}{binary phase-shift keying}
  \acro{BRA}{balanced random allocation}
  \acro{BS}{base station}
  \acro{BSUM}{block successive upper-bound minimization}
  \acro{CAP}{combinatorial allocation problem}
  \acro{CAPEX}{capital expenditure}
  \acro{CBF}{coordinated beamforming}
  \acro{CBR}{constant bit rate}
  \acro{CBS}{class based scheduling}
  \acro{CC}{congestion control}
  \acro{CDF}{cumulative distribution function}
  \acro{CDMA}{code-division multiple access}
  \acro{CE}{\LU{C}{c}hannel \LU{E}{e}stimation}
  \acro{CL}{closed loop}
  \acro{CLPC}{closed loop power control}
  \acro{CML}{centralized machine learning}
  \acro{CNR}{channel-to-noise ratio}
  \acro{CNN}{\LU{C}{c}onvolutional \LU{N}{n}eural \LU{N}{n}etwork}
  \acro{CP}{computation point}
  \acro{CPA}{cellular protection algorithm}
  \acro{CPICH}{common pilot channel}
  \acro{CoCoA}{\LU{C}{c}ommunication efficient distributed dual \LU{C}{c}oordinate \LU{A}{a}scent}
  \acro{CoMAC}{\LU{C}{c}omputation over \LU{M}{m}ultiple-\LU{A}{a}ccess \LU{C}{c}hannels}
  \acro{CoMP}{coordinated multi-point}
  \acro{Comp}{Summation Computing}
  \acro{CQI}{channel quality indicator}
  \acro{CRM}{constrained rate maximization}
  \acro{CRN}{cognitive radio network}
  \acro{CS}{coordinated scheduling}
  \acro{CSI}{\LU{C}{c}hannel \LU{S}{s}tate \LU{I}{i}nformation}
  \acro{CSMA}{\LU{C}{c}arrier \LU{S}{s}ense \LU{M}{m}ultiple \LU{A}{a}ccess}
  \acro{CUE}{cellular user equipment}
  \acro{D2D}{device-to-device}
  \acro{DAC}{digital-to-analog converter}
  \acro{DC}{direct current}
  \acro{DCA}{dynamic channel allocation}
  \acro{DE}{differential evolution}
  \acro{DFT}{discrete Fourier transform}
  \acro{DIST}{distance}
  \acro{DL}{downlink}
  \acro{DMA}{double moving average}
  \acro{DML}{distributed machine learning}
  \acro{DMRS}{demodulation reference signal}
  \acro{D2DM}{D2D mode}
  \acro{DMS}{D2D mode selection}
  \acro{DNN}{deep neural network}
  \acro{DPC}{dirty paper coding}
  \acro{DRA}{dynamic resource assignment}
  \acro{DSA}{dynamic spectrum access}
  \acro{DSGD}{\LU{D}{d}istributed \LU{S}{s}tochastic \LU{G}{g}radient \LU{D}{d}escent}
  \acro{DSM}{delay-based satisfaction maximization}
  \acro{ECC}{electronic communications committee}
  \acro{EFLC}{error feedback based load control}
  \acro{EI}{efficiency indicator}
  \acro{eNB}{evolved Node B}
  \acro{EPA}{equal power allocation}
  \acro{EPC}{evolved packet core}
  \acro{EPS}{evolved packet system}
  \acro{E-UTRAN}{evolved universal terrestrial radio access network}
  \acro{ES}{edge server}
  \acro{ED}{edge device}
  \acro{FC}{\LU{F}{f}usion \LU{C}{c}enter}
  \acro{FSK}{frequency-shift keying}
  \acro{FD}{\LU{F}{f}ederated \LU{D}{d}istillation}
  \acro{FDD}{frequency division duplex}
  \acro{FDM}{frequency division multiplexing}
  \acro{FDMA}{\LU{F}{f}requency \LU{D}{d}ivision \LU{M}{m}ultiple \LU{A}{a}ccess}
  \acro{FedAvg}{\LU{F}{f}ederated \LU{A}{a}veraging}
  \acro{FER}{frame erasure rate}
  \acro{FF}{fast fading}
  \acro{FL}{federated learning}
  \acro{FEEL}{federated edge learning}
  \acro{FSB}{fixed switched beamforming}
  \acro{FST}{fixed SNR target}
  \acro{FTP}{file transfer protocol}
  \acro{GA}{genetic algorithm}
  \acro{GBR}{guaranteed bit rate}
  \acro{GD}{gradient descent}
  \acro{GLR}{gain to leakage ratio}
  \acro{GOS}{generated orthogonal sequence}
  \acro{GPL}{GNU general public license}
  \acro{GRP}{grouping}
  \acro{HARQ}{hybrid automatic repeat request}
  \acro{HD}{half-duplex}
  \acro{HMS}{harmonic mode selection}
  \acro{HOL}{head of line}
  \acro{HSDPA}{high-speed downlink packet access}
  \acro{HSPA}{high speed packet access}
  \acro{HTTP}{hypertext transfer protocol}
  \acro{ICMP}{internet control message protocol}
  \acro{ICI}{intercell interference}
  \acro{ID}{identification}
  \acro{IETF}{internet engineering task force}
  \acro{ILP}{integer linear program}
  \acro{JRAPAP}{joint RB assignment and power allocation problem}
  \acro{UID}{unique identification}
  \acro{iid}{independent and identically distributed}
  \acro{IIR}{infinite impulse response}
  \acro{ILP}{integer linear problem}
  \acro{IMT}{international mobile telecommunications}
  \acro{INV}{inverted norm-based grouping}
  \acro{IoT}{internet of things}
  \acro{IP}{integer programming}
  \acro{IPv6}{internet protocol version 6}
  \acro{IQ}{in-phase quadrature}
  \acro{ISD}{inter-site distance}
  \acro{ISI}{inter symbol interference}
  \acro{ITU}{international telecommunication union}
  \acro{JAFM}{joint assignment and fairness maximization}
  \acro{JAFMA}{joint assignment and fairness maximization algorithm}
  \acro{JOAS}{joint opportunistic assignment and scheduling}
  \acro{JOS}{joint opportunistic scheduling}
  \acro{JP}{joint processing}
	\acro{JS}{jump-stay}
  \acro{KKT}{Karush-Kuhn-Tucker}
  \acro{L3}{Layer-3}
  \acro{LAC}{link admission control}
  \acro{LA}{link adaptation}
  \acro{LC}{load control}
  \acro{LDC}{\LU{L}{l}earning-\LU{D}{d}riven \LU{C}{c}ommunication}
  \acro{LOS}{line of sight}
  \acro{LP}{linear programming}
  \acro{LTE}{long term evolution}
	\acro{LTE-A}{\ac{LTE}-advanced}
  \acro{LTE-Advanced}{long term evolution advanced}
  \acro{M2M}{machine-to-machine}
  \acro{MAC}{multiple access channel}
  \acro{MANET}{mobile ad hoc network}
  \acro{MC}{modular clock}
  \acro{MCS}{modulation and coding scheme}
  \acro{MDB}{measured delay based}
  \acro{MDI}{minimum D2D interference}
  \acro{MF}{matched filter}
  \acro{MG}{maximum gain}
  \acro{MH}{multi-hop}
  \acro{MIMO}{\LU{M}{m}ultiple \LU{I}{i}nput \LU{M}{m}ultiple \LU{O}{o}utput}
  \acro{MINLP}{mixed integer nonlinear programming}
  \acro{MIP}{mixed integer programming}
  \acro{MISO}{multiple input single output}
  \acro{ML}{machine learning}
  \acro{MLWDF}{modified largest weighted delay first}
  \acro{MME}{mobility management entity}
  \acro{MMSE}{minimum mean squared error}
  \acro{MOS}{mean opinion score}
  \acro{MPF}{multicarrier proportional fair}
  \acro{MRA}{maximum rate allocation}
  \acro{MR}{maximum rate}
  \acro{MRC}{maximum ratio combining}
  \acro{MRT}{maximum ratio transmission}
  \acro{MRUS}{maximum rate with user satisfaction}
  \acro{MS}{mode selection}
  \acro{MAE}{\LU{M}{m}ean \LU{A}{a}bsolute \LU{E}{e}rror}
  \acro{MSE}{\LU{M}{m}ean \LU{S}{s}quared \LU{E}{e}rror}
  \acro{MSI}{multi-stream interference}
  \acro{MTC}{machine-type communication}
  \acro{MTSI}{multimedia telephony services over IMS}
  \acro{MTSM}{modified throughput-based satisfaction maximization}
  \acro{MU-MIMO}{multi-user multiple input multiple output}
  \acro{MU}{multi-user}
  \acro{NAS}{non-access stratum}
  \acro{NB}{Node B}
  \acro{NCL}{neighbor cell list}
  \acro{NLP}{nonlinear programming}
  \acro{NLOS}{non-line of sight}
  \acro{NMSE}{normalized mean square error}
  \acro{NN}{neural network}
  \acro{NOMA}{\LU{N}{n}on-\LU{O}{o}rthogonal \LU{M}{m}ultiple \LU{A}{a}ccess}
  \acro{NORM}{normalized projection-based grouping}
  \acro{NP}{non-polynomial time}
  \acro{NRT}{non-real time}
  \acro{NSPS}{national security and public safety services}
  \acro{O2I}{outdoor to indoor}
  \acro{OAC}{over-the-air computation}
  \acro{OFDMA}{\LU{O}{o}rthogonal \LU{F}{f}requency \LU{D}{d}ivision \LU{M}{m}ultiple \LU{A}{a}ccess}
  \acro{OFDM}{orthogonal frequency division multiplexing}
  \acro{OFPC}{open loop with fractional path loss compensation}
	\acro{O2I}{outdoor-to-indoor}
  \acro{OL}{open loop}
  \acro{OLPC}{open-loop power control}
  \acro{OL-PC}{open-loop power control}
  \acro{OPEX}{operational expenditure}
  \acro{ORB}{orthogonal random beamforming}
  \acro{JO-PF}{joint opportunistic proportional fair}
  \acro{OSI}{open systems interconnection}
  \acro{PM}{phase modulation}
  \acro{PAM}{pulse-amplitude modulation}
  \acro{PAIR}{D2D pair gain-based grouping}
  \acro{PAPR}{peak-to-average power ratio}
  \acro{P2P}{peer-to-peer}
  \acro{PC}{power control}
  \acro{PCI}{physical cell ID}
  \acro{PDCCH}{physical downlink control channel}
  \acro{PDD}{penalty dual decomposition}
  \acro{PDF}{probability density function}
  \acro{PER}{packet error rate}
  \acro{PF}{proportional fair}
  \acro{P-GW}{packet data network gateway}
  \acro{PL}{pathloss}
  \acro{PLL}{phase-locked loop}
  \acro{PRB}{physical resource block}
  \acro{PROJ}{projection-based grouping}
  \acro{ProSe}{proximity services}
  \acro{PS}{\LU{P}{p}arameter \LU{S}{s}erver}
  \acro{PSO}{particle swarm optimization}
  \acro{PUCCH}{physical uplink control channel}
  \acro{PZF}{projected zero-forcing}
  \acro{QAM}{quadrature amplitude modulation}
  \acro{QoS}{quality of service}
  \acro{QPSK}{quadri-phase shift keying}
  \acro{RAISES}{reallocation-based assignment for improved spectral efficiency and satisfaction}
  \acro{RAN}{radio access network}
  \acro{RA}{resource allocation}
  \acro{RAT}{radio access technology}
  \acro{RATE}{rate-based}
  \acro{RB}{resource block}
  \acro{RBG}{resource block group}
  \acro{REF}{reference grouping}
  \acro{RF}{radio frequency}
  \acro{RLC}{radio link control}
  \acro{RM}{rate maximization}
  \acro{RNC}{radio network controller}
  \acro{RND}{random grouping}
  \acro{RRA}{radio resource allocation}
  \acro{RRM}{\LU{R}{r}adio \LU{R}{r}esource \LU{M}{m}anagement}
  \acro{RSCP}{received signal code power}
  \acro{RSRP}{reference signal receive power}
  \acro{RSRQ}{reference signal receive quality}
  \acro{RR}{round robin}
  \acro{RRC}{radio resource control}
  \acro{RSSI}{received signal strength indicator}
  \acro{RT}{real time}
  \acro{RU}{resource unit}
  \acro{RUNE}{rudimentary network emulator}
  \acro{RV}{random variable}
  \acro{SAC}{session admission control}
  \acro{SCM}{spatial channel model}
  \acro{SC-FDMA}{single carrier - frequency division multiple access}
  \acro{SD}{soft dropping}
  \acro{S-D}{source-destination}
  \acro{SDPC}{soft dropping power control}
  \acro{SDMA}{space-division multiple access}
  \acro{SDR}{semidefinite relaxation}
  \acro{SDP}{semidefinite programming}
  \acro{SER}{symbol error rate}
  \acro{SES}{simple exponential smoothing}
  \acro{S-GW}{serving gateway}
  \acro{SGD}{\LU{S}{s}tochastic \LU{G}{g}radient \LU{D}{d}escent}  
  \acro{SINR}{signal-to-interference-plus-noise ratio}
  \acro{SI}{self-interference}
  \acro{SIP}{Session Initiation Protocol}
  \acro{SISO}{\LU{S}{s}ingle \LU{I}{i}nput \LU{S}{s}ingle \LU{O}{o}utput}
  \acro{SIMO}{Single Input Multiple Output}
  \acro{SIR}{signal to interference ratio}
  \acro{SLNR}{Signal-to-Leakage-plus-Noise Ratio}
  \acro{SMA}{simple moving average}
  \acro{SNR}{\LU{S}{s}ignal-to-\LU{N}{n}oise \LU{R}{r}atio}
  \acro{SORA}{satisfaction oriented resource allocation}
  \acro{SORA-NRT}{satisfaction-oriented resource allocation for non-real time services}
  \acro{SORA-RT}{satisfaction-oriented resource allocation for real time services}
  \acro{SPF}{single-carrier proportional fair}
  \acro{SRA}{sequential removal algorithm}
  \acro{SRS}{sounding reference signal}
  \acro{SU-MIMO}{single-user multiple input multiple output}
  \acro{SU}{single-user}
  \acro{SVD}{singular value decomposition}
  \acro{SVM}{\LU{S}{s}upport \LU{V}{v}ector \LU{M}{m}achine}
  \acro{TCP}{Transmission Control Protocol}
  \acro{TDD}{time division duplex}
  \acro{TDMA}{\LU{T}{t}ime \LU{D}{d}ivision \LU{M}{m}ultiple \LU{A}{a}ccess}
  \acro{TNFD}{three node full duplex}
  \acro{TETRA}{terrestrial trunked radio}
  \acro{TP}{transmit power}
  \acro{TPC}{transmit power control}
  \acro{TTI}{transmission time interval}
  \acro{TTR}{time-to-rendezvous}
  \acro{TSM}{throughput-based satisfaction maximization}
  \acro{TU}{typical urban}
  \acro{UE}{\LU{U}{u}ser \LU{E}{e}quipment}
  \acro{UEPS}{urgency and efficiency-based packet scheduling}
  \acro{UL}{uplink}
  \acro{UMTS}{universal mobile telecommunications system}
  \acro{URI}{uniform resource identifier}
  \acro{URM}{unconstrained rate maximization}
  \acro{VR}{virtual resource}
  \acro{VoIP}{voice over IP}
  \acro{WAN}{wireless access network}
  \acro{WCDMA}{wideband code division multiple access}
  \acro{WF}{water-filling}
  \acro{WiMAX}{worldwide interoperability for microwave access}
  \acro{WINNER}{wireless world initiative new radio}
  \acro{WLAN}{wireless local area network}
  \acro{WMMSE}{weighted minimum mean square error}
  \acro{WMPF}{weighted multicarrier proportional fair}
  \acro{WPF}{weighted proportional fair}
  \acro{WSN}{wireless sensor network}
  \acro{WWW}{world wide web}
  \acro{XIXO}{(single or multiple) input (single or multiple) output}
  \acro{ZF}{zero-forcing}
  \acro{ZMCSCG}{zero mean circularly symmetric complex Gaussian}
\end{acronym}

%% file: fig/Fig_Feel.tex
\begin{figure*}[!t]
    \centering
\scalebox{0.8}{
\tikzset{every picture/.style={line width=0.75pt}} 

\begin{tikzpicture}[x=0.75pt,y=0.75pt,yscale=-1,xscale=1]

\draw  [fill={rgb, 255:red, 40; green, 33; blue, 33 }  ,fill opacity=1 ] (553.21,114.23) .. controls (553.21,109.93) and (562.06,106.45) .. (572.99,106.45) .. controls (583.91,106.45) and (592.77,109.93) .. (592.77,114.23) .. controls (592.77,118.52) and (583.91,122) .. (572.99,122) .. controls (562.06,122) and (553.21,118.52) .. (553.21,114.23) -- cycle ;
\draw  [color={rgb, 255:red, 255; green, 252; blue, 252 }  ,draw opacity=1 ][fill={rgb, 255:red, 247; green, 246; blue, 246 }  ,fill opacity=1 ] (553.21,110.34) .. controls (553.21,106.05) and (561.92,102.57) .. (572.66,102.57) .. controls (583.4,102.57) and (592.11,106.05) .. (592.11,110.34) .. controls (592.11,114.63) and (583.4,118.11) .. (572.66,118.11) .. controls (561.92,118.11) and (553.21,114.63) .. (553.21,110.34) -- cycle ;
\draw  [fill={rgb, 255:red, 40; green, 33; blue, 33 }  ,fill opacity=1 ] (553.21,106.45) .. controls (553.21,102.16) and (562.06,98.68) .. (572.99,98.68) .. controls (583.91,98.68) and (592.77,102.16) .. (592.77,106.45) .. controls (592.77,110.75) and (583.91,114.23) .. (572.99,114.23) .. controls (562.06,114.23) and (553.21,110.75) .. (553.21,106.45) -- cycle ;
\draw  [color={rgb, 255:red, 255; green, 252; blue, 252 }  ,draw opacity=1 ][fill={rgb, 255:red, 247; green, 246; blue, 246 }  ,fill opacity=1 ] (553.21,102.57) .. controls (553.21,98.28) and (562.06,94.8) .. (572.99,94.8) .. controls (583.91,94.8) and (592.77,98.28) .. (592.77,102.57) .. controls (592.77,106.86) and (583.91,110.34) .. (572.99,110.34) .. controls (562.06,110.34) and (553.21,106.86) .. (553.21,102.57) -- cycle ;
\draw  [fill={rgb, 255:red, 40; green, 33; blue, 33 }  ,fill opacity=1 ] (553.21,98.68) .. controls (553.21,94.39) and (562.06,90.91) .. (572.99,90.91) .. controls (583.91,90.91) and (592.77,94.39) .. (592.77,98.68) .. controls (592.77,102.97) and (583.91,106.45) .. (572.99,106.45) .. controls (562.06,106.45) and (553.21,102.97) .. (553.21,98.68) -- cycle ;
\draw  [color={rgb, 255:red, 255; green, 252; blue, 252 }  ,draw opacity=1 ][fill={rgb, 255:red, 247; green, 246; blue, 246 }  ,fill opacity=1 ] (553.21,94.8) .. controls (553.21,90.5) and (562.21,87.02) .. (573.32,87.02) .. controls (584.43,87.02) and (593.43,90.5) .. (593.43,94.8) .. controls (593.43,99.09) and (584.43,102.57) .. (573.32,102.57) .. controls (562.21,102.57) and (553.21,99.09) .. (553.21,94.8) -- cycle ;
\draw  [fill={rgb, 255:red, 40; green, 33; blue, 33 }  ,fill opacity=1 ] (553.21,90.91) .. controls (553.21,86.62) and (562.06,83.14) .. (572.99,83.14) .. controls (583.91,83.14) and (592.77,86.62) .. (592.77,90.91) .. controls (592.77,95.2) and (583.91,98.68) .. (572.99,98.68) .. controls (562.06,98.68) and (553.21,95.2) .. (553.21,90.91) -- cycle ;

\draw    (552,103) -- (520,103) ;
\draw [shift={(520,103)}, rotate = 360] [fill={rgb, 255:red, 0; green, 0; blue, 0 }  ][line width=0.08]  [draw opacity=0] (3.57,-1.72) -- (0,0) -- (3.57,1.72) -- cycle    ;
\draw    (405,103) -- (387,103) ;
\draw [shift={(387,103)}, rotate = 360] [fill={rgb, 255:red, 0; green, 0; blue, 0 }  ][line width=0.08]  [draw opacity=0] (3.57,-1.72) -- (0,0) -- (3.57,1.72) -- cycle    ;
\draw    (298,103) -- (271,103) ;
\draw [shift={(271,103)}, rotate = 360] [fill={rgb, 255:red, 0; green, 0; blue, 0 }  ][line width=0.08]  [draw opacity=0] (3.57,-1.72) -- (0,0) -- (3.57,1.72) -- cycle    ;
\draw    (534,57) -- (534,99) ;
\draw [shift={(534,102)}, rotate = 270] [fill={rgb, 255:red, 0; green, 0; blue, 0 }  ][line width=0.08]  [draw opacity=0] (3.57,-1.72) -- (0,0) -- (3.57,1.72) -- cycle    ;
\draw    (480,57) -- (534,57) ;
\draw   (263,85) -- (256.11,76) -- (270,76.12) -- cycle ;
\draw    (263,85) -- (263,105) ;
\draw  [draw opacity=1][fill={rgb, 1:red, 0.6; green, 0.47; blue, 0.48 }  ,fill opacity=0.7 ] (259, 105) -- (269,105) -- (269,115) -- (259,115) -- cycle ;

\draw  [fill={rgb, 255:red, 40; green, 33; blue, 33 }  ,fill opacity=1 ] (554.21,251.23) .. controls (554.21,246.93) and (563.06,243.45) .. (573.99,243.45) .. controls (584.91,243.45) and (593.77,246.93) .. (593.77,251.23) .. controls (593.77,255.52) and (584.91,259) .. (573.99,259) .. controls (563.06,259) and (554.21,255.52) .. (554.21,251.23) -- cycle ;
\draw  [color={rgb, 255:red, 255; green, 252; blue, 252 }  ,draw opacity=1 ][fill={rgb, 255:red, 247; green, 246; blue, 246 }  ,fill opacity=1 ] (554.21,247.34) .. controls (554.21,243.05) and (562.92,239.57) .. (573.66,239.57) .. controls (584.4,239.57) and (593.11,243.05) .. (593.11,247.34) .. controls (593.11,251.63) and (584.4,255.11) .. (573.66,255.11) .. controls (562.92,255.11) and (554.21,251.63) .. (554.21,247.34) -- cycle ;
\draw  [fill={rgb, 255:red, 40; green, 33; blue, 33 }  ,fill opacity=1 ] (554.21,243.45) .. controls (554.21,239.16) and (563.06,235.68) .. (573.99,235.68) .. controls (584.91,235.68) and (593.77,239.16) .. (593.77,243.45) .. controls (593.77,247.75) and (584.91,251.23) .. (573.99,251.23) .. controls (563.06,251.23) and (554.21,247.75) .. (554.21,243.45) -- cycle ;
\draw  [color={rgb, 255:red, 255; green, 252; blue, 252 }  ,draw opacity=1 ][fill={rgb, 255:red, 247; green, 246; blue, 246 }  ,fill opacity=1 ] (554.21,239.57) .. controls (554.21,235.28) and (563.06,231.8) .. (573.99,231.8) .. controls (584.91,231.8) and (593.77,235.28) .. (593.77,239.57) .. controls (593.77,243.86) and (584.91,247.34) .. (573.99,247.34) .. controls (563.06,247.34) and (554.21,243.86) .. (554.21,239.57) -- cycle ;
\draw  [fill={rgb, 255:red, 40; green, 33; blue, 33 }  ,fill opacity=1 ] (554.21,235.68) .. controls (554.21,231.39) and (563.06,227.91) .. (573.99,227.91) .. controls (584.91,227.91) and (593.77,231.39) .. (593.77,235.68) .. controls (593.77,239.97) and (584.91,243.45) .. (573.99,243.45) .. controls (563.06,243.45) and (554.21,239.97) .. (554.21,235.68) -- cycle ;
\draw  [color={rgb, 255:red, 255; green, 252; blue, 252 }  ,draw opacity=1 ][fill={rgb, 255:red, 247; green, 246; blue, 246 }  ,fill opacity=1 ] (554.21,231.8) .. controls (554.21,227.5) and (563.21,224.02) .. (574.32,224.02) .. controls (585.43,224.02) and (594.43,227.5) .. (594.43,231.8) .. controls (594.43,236.09) and (585.43,239.57) .. (574.32,239.57) .. controls (563.21,239.57) and (554.21,236.09) .. (554.21,231.8) -- cycle ;
\draw  [fill={rgb, 255:red, 40; green, 33; blue, 33 }  ,fill opacity=1 ] (554.21,227.91) .. controls (554.21,223.62) and (563.06,220.14) .. (573.99,220.14) .. controls (584.91,220.14) and (593.77,223.62) .. (593.77,227.91) .. controls (593.77,232.2) and (584.91,235.68) .. (573.99,235.68) .. controls (563.06,235.68) and (554.21,232.2) .. (554.21,227.91) -- cycle ;

\draw    (553,239) -- (520,239) ;
\draw [shift={(520,239)}, rotate = 360] [fill={rgb, 255:red, 0; green, 0; blue, 0 }  ][line width=0.08]  [draw opacity=0] (3.57,-1.72) -- (0,0) -- (3.57,1.72) -- cycle    ;
\draw    (405,240) -- (387,240) ;
\draw [shift={(387,240)}, rotate = 0.5] [fill={rgb, 255:red, 0; green, 0; blue, 0 }  ][line width=0.08]  [draw opacity=0] (3.57,-1.72) -- (0,0) -- (3.57,1.72) -- cycle    ;
\draw    (298,240) -- (275,240) ;
\draw [shift={(272,240)}, rotate = 0.38] [fill={rgb, 255:red, 0; green, 0; blue, 0 }  ][line width=0.08]  [draw opacity=0] (3.57,-1.72) -- (0,0) -- (3.57,1.72) -- cycle    ;
\draw    (535,195) -- (535,236) ;
\draw [shift={(535,239)}, rotate = 270] [fill={rgb, 255:red, 0; green, 0; blue, 0 }  ][line width=0.08]  [draw opacity=0] (3.57,-1.72) -- (0,0) -- (3.57,1.72) -- cycle    ;
\draw    (480,195) -- (535,195) ;
\draw   (263.96,221.97) -- (257.11,213) -- (271,213.12) -- cycle ;
\draw    (263.96,221.97) -- (264,245) ;
\draw  [draw opacity=1][fill={rgb, 1:red, 0.6; green, 0.47; blue, 0.48 }  ,fill opacity=0.7 ] (259, 245) -- (269,245) -- (269,255) -- (259,255) -- cycle ;

\draw [line width=1.5]    (158.72,122.99) -- (140,182.45) ;
\draw [line width=1.5]    (158.72,122.99) -- (161.39,185) ;
\draw [line width=1.5]    (158.72,122.99) -- (181,178.2) ;
\draw    (160.05,153.99) -- (169.86,150.6) ;
\draw    (173.87,159.8) -- (160.5,165.46) ;
\draw    (165.85,140.26) -- (159.61,142.53) ;
\draw    (177.43,169.14) -- (161.39,175.66) ;
\draw    (161.39,175.66) -- (143.57,169.99) ;
\draw    (160.5,165.46) -- (146.24,161.5) ;
\draw    (149.36,152.72) -- (160.05,153.99) ;
\draw    (152.48,141.11) -- (159.61,142.53) ;
\draw  [draw opacity=0][fill={rgb, 255:red, 156; green, 133; blue, 133 }  ,fill opacity=1 ] (145.35,106) -- (172.09,106) -- (172.09,123.83) -- (145.35,123.83) -- cycle ; \draw   (149.8,106) -- (149.8,123.83)(154.26,106) -- (154.26,123.83)(158.72,106) -- (158.72,123.83)(163.17,106) -- (163.17,123.83)(167.63,106) -- (167.63,123.83) ; \draw   (145.35,110.46) -- (172.09,110.46)(145.35,114.91) -- (172.09,114.91)(145.35,119.37) -- (172.09,119.37) ; \draw   (145.35,106) -- (172.09,106) -- (172.09,123.83) -- (145.35,123.83) -- cycle ;

\draw   [color={rgb, 255:red, 190; green, 50; blue, 20 }  ] (251,98) -- (191.45,140.26)   ;
\draw [shift={(189,142)}, rotate = 324.64] [fill={rgb, 255:red, 190; green, 50; blue, 20  }  ][line width=0.08]  [draw opacity=0] (3.57,-1.72) -- (0,0) -- (3.57,1.72) -- cycle    ;

\draw [color={rgb, 255:red, 20; green, 120; blue, 190 }  ] [line width=0.75]  [dash pattern={on 0.84pt off 2.51pt}]  (252.55,105.74) -- (226,124.58) -- (193,148) ;
\draw [shift={(255,104)}, rotate = 144.64] [fill={rgb, 255:red, 20; green, 120; blue, 190 }  ][line width=0.08]  [draw opacity=0] (3.57,-1.72) -- (0,0) -- (3.57,1.72) -- cycle    ;

\draw   [color={rgb, 255:red, 190; green, 50; blue, 20 }  ] (254.91,223.56) -- (195.93,180.51) ;
\draw [shift={(193.51,178.74)}, rotate = 36.13] [fill={rgb, 255:red, 190; green, 50; blue, 20 }  ][line width=0.08]  [draw opacity=0] (3.57,-1.72) -- (0,0) -- (3.57,1.72) -- cycle    ;
\draw [color={rgb, 255:red, 20; green, 120; blue, 190 }  ] [line width=0.75]  [dash pattern={on 0.84pt off 2.51pt}]  (248.07,227.49) -- (221.77,208.29) -- (189.09,184.44) ;
\draw [shift={(250.49,229.26)}, rotate = 216.13] [fill={rgb, 255:red, 20; green, 120; blue, 190 }  ][line width=0.08]  [draw opacity=0] (3.57,-1.72) -- (0,0) -- (3.57,1.72) -- cycle    ;

\draw    (286,57) -- (328,57) ;
\draw [shift={(328,57)}, rotate = 180] [fill={rgb, 255:red, 0; green, 0; blue, 0 }  ][line width=0.08]  [draw opacity=0] (3.57,-1.72) -- (0,0) -- (3.57,1.72) -- cycle    ;
\draw    (286,57) -- (270,71) ;

\draw    (80,164) -- (80,207) ;
\draw    (80,207) -- (154,207) ;
\draw    (154,190) -- (154,207) ;

\draw [shift={(154,187)}, rotate = 90] [fill={rgb, 255:red, 0; green, 0; blue, 0 }  ][line width=0.08]  [draw opacity=0] (3.57,-1.72) -- (0,0) -- (3.57,1.72) -- cycle    ;
\draw    (287,195) -- (328,195) ;
\draw [shift={(328,195)}, rotate = 180] [fill={rgb, 255:red, 0; green, 0; blue, 0 }  ][line width=0.08]  [draw opacity=0] (3.57,-1.72) -- (0,0) -- (3.57,1.72) -- cycle    ;
\draw    (287,195) -- (271,208) ;

\draw    (405,89) -- (520,89) -- (520,115) -- (405,115) -- cycle  ;
\draw (407,93.4) node [anchor=north west][inner sep=0.75pt]  [font=\footnotesize]  {$\bm{g}_{1}{(m)} = \nabla \mathcal{L}_{1}(\bm{w}{(m)})$};
\draw    (298,88) -- (387,88) -- (387,117) -- (298,117) -- cycle  ;
\draw (300,92) node [anchor=north west][inner sep=0.75pt]  [font=\footnotesize]  {$\bm{s}_1=\bm{g}_{1}{(m)}$};
\draw    (328,47) -- (480,47) -- (480,69) -- (328,69) -- cycle  ;
\draw (331,51.4) node [anchor=north west][inner sep=0.75pt]  [font=\footnotesize]  {$\hspace{-2pt}\bm{w}{(m)}\hspace{-1.5pt} = \hspace{-1.5pt}\bm{w}{(m-1)}\hspace{-2pt} -\eta \bm{g}{(m-1)}$};
\draw (545,60) node [anchor=north west][inner sep=0.75pt]   [align=left] {{\footnotesize Local Dataset $\mathcal{D}_1$ }};
\draw    (405,225) -- (520,225) -- (520,251) -- (405,251) -- cycle  ;
\draw (407,229) node [anchor=north west][inner sep=0.75pt]  [font=\footnotesize]  {$\bm{g}_{K}{(m)}  = \nabla \mathcal{L}_{K}\hspace{-1.5pt}(w{(m)})$};
\draw    (298,224) -- (387,224) -- (387,252) -- (298,252) -- cycle  ;
\draw (302,228.4) node [anchor=north west][inner sep=0.75pt]  [font=\footnotesize]  {$\bm{s}_K=  \bm{g}_K(m)$};
\draw    (328,184) -- (480,184) -- (480,206) -- (328,206) -- cycle  ;
\draw (332,188.4) node [anchor=north west][inner sep=0.75pt]  [font=\footnotesize]  {$\hspace{-3pt}\bm{w}{(m)}\hspace{-1pt} =\hspace{-1pt}\bm{w}{(m-1)} \hspace{-2.5pt}-\eta {\bm{g}}{(m-1)}$};
\draw (545,196) node [anchor=north west][inner sep=0.75pt]   [align=left] {{\footnotesize Local Dataset  $\mathcal{D}_K$}};
\draw (565,130) node [anchor=north west][inner sep=0.75pt]  [font=\LARGE]  {$\vdots $};
\draw (390,132) node [anchor=north west][inner sep=0.75pt]  [font=\LARGE]  {$\vdots $};
\draw    (10,136) -- (138,136) -- (138,158) -- (10,158) -- cycle  ;
\draw (77,147) node  [font=\footnotesize]  {$\hspace{-3pt}\bm{g}{(m)}= \sum_{k}\bm{g}_k{(m)}/K$};
\draw (115,220) node  [font=\footnotesize]  {$ \bm{w}{(m+1)} = \bm{w}{(m)} - \eta \bm{g}{(m)}$};

\end{tikzpicture}
}
    \caption{Diagram of federated edge learning where we assume to perform over-the-air computation by the digital modulation QAM. Here, dashed arrow lines (blue color) show the downlink transmission, where the \ac{ES} sends the updated global model in Eq.~\eqref{eq:aggreg} back to the edge device. The red lines show the uplink phase, where device $k$ transmits the parameters of the trained model with its local data using modulated signal $\bm{s}_k$ in Eq.~\eqref{eq:Sk} to the server \ac{ES}.   }
    \label{fig:federated}
\end{figure*}
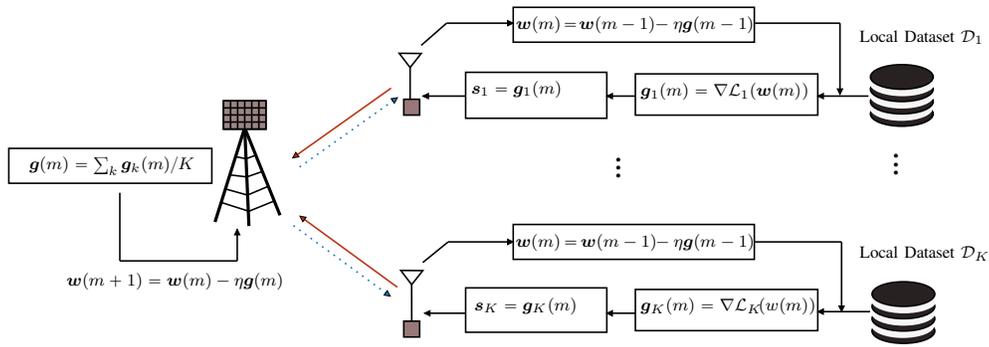

%% file: fig/FigSumCode.tex
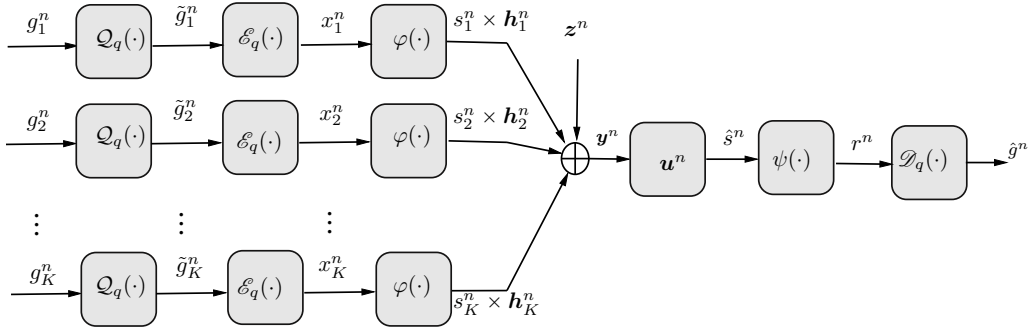
\begin{figure*}
    \centering

\scalebox{0.85}{

\tikzset{every picture/.style={line width=0.75pt}} 

\begin{tikzpicture}[x=0.75pt,y=0.75pt,yscale=-1,xscale=1]

\draw   (351.35,108.65) .. controls (351.35,103.48) and (354.99,99.29) .. (359.47,99.29) .. controls (363.96,99.29) and (367.6,103.48) .. (367.6,108.65) .. controls (367.6,113.82) and (363.96,118.01) .. (359.47,118.01) .. controls (354.99,118.01) and (351.35,113.82) .. (351.35,108.65) -- cycle ; \draw   (351.35,108.65) -- (367.6,108.65) ; \draw   (359.47,99.29) -- (359.47,118.01) ;
\draw [line width=0.75]    (319,187) -- (354.22,119.31) ;
\draw [shift={(355.07,117.5)}, rotate = 115.03] [fill={rgb, 255:red, 0; green, 0; blue, 0 }  ][line width=0.08]  [draw opacity=0] (7.2,-1.8) -- (0,0) -- (7.2,1.8) -- cycle    ;
\draw  [color={rgb, 255:red, 21; green, 20; blue, 20 }  ,draw opacity=1 ][fill={rgb, 255:red, 230; green, 230; blue, 230  }  ,fill opacity=1 ] (547.37,97.97) .. controls (547.37,93.15) and (551.28,89.24) .. (556.1,89.24) -- (583.11,89.24) .. controls (587.93,89.24) and (591.84,93.15) .. (591.84,97.97) -- (591.84,124.16) .. controls (591.84,128.98) and (587.93,132.89) .. (583.11,132.89) -- (556.1,132.89) .. controls (551.28,132.89) and (547.37,128.98) .. (547.37,124.16) -- cycle ;
\draw [line width=0.75]    (319,99) -- (349.13,105.06) ;
\draw [shift={(351.09,105.46)}, rotate = 191.55] [fill={rgb, 255:red, 0; green, 0; blue, 0 }  ][line width=0.08]  [draw opacity=0] (7.2,-1.8) -- (0,0) -- (7.2,1.8) -- cycle    ;
\draw [line width=0.75]    (367.01,108.65) -- (389.5,108.97) ;
\draw [shift={(391.5,109)}, rotate = 180.83] [fill={rgb, 255:red, 0; green, 0; blue, 0 }  ][line width=0.08]  [draw opacity=0] (7.2,-1.8) -- (0,0) -- (7.2,1.8) -- cycle    ;
\draw [line width=0.75]    (514.25,111.55) -- (545.5,111.97) ;
\draw [shift={(547.5,112)}, rotate = 180.77] [fill={rgb, 255:red, 0; green, 0; blue, 0 }  ][line width=0.08]  [draw opacity=0] (7.2,-1.8) -- (0,0) -- (7.2,1.8) -- cycle    ;
\draw [line width=0.75]    (361.04,49.63) -- (360.44,90.67) -- (360.44,95.23) ;
\draw [shift={(360.44,97.23)}, rotate = 270] [fill={rgb, 255:red, 0; green, 0; blue, 0 }  ][line width=0.08]  [draw opacity=0] (7.2,-1.8) -- (0,0) -- (7.2,1.8) -- cycle    ;
\draw [line width=0.75]    (591.27,110.99) -- (613.5,111) ;
\draw [shift={(615.5,111)}, rotate = 180.02] [fill={rgb, 255:red, 0; green, 0; blue, 0 }  ][line width=0.08]  [draw opacity=0] (7.2,-1.8) -- (0,0) -- (7.2,1.8) -- cycle    ;
\draw  [color={rgb, 255:red, 21; green, 20; blue, 20 }  ,draw opacity=1 ][fill={rgb, 255:red, 230; green, 230; blue, 230  }  ,fill opacity=1 ] (150.46,83.03) .. controls (150.46,78.2) and (154.37,74.3) .. (159.19,74.3) -- (186.2,74.3) .. controls (191.02,74.3) and (194.93,78.2) .. (194.93,83.03) -- (194.93,109.21) .. controls (194.93,114.03) and (191.02,117.94) .. (186.2,117.94) -- (159.19,117.94) .. controls (154.37,117.94) and (150.46,114.03) .. (150.46,109.21) -- cycle ;
\draw  [color={rgb, 255:red, 21; green, 20; blue, 20 }  ,draw opacity=1 ][fill={rgb, 255:red, 230; green, 230; blue, 230  }  ,fill opacity=1 ] (153.45,172.09) .. controls (153.45,167.27) and (157.36,163.36) .. (162.18,163.36) -- (189.19,163.36) .. controls (194.01,163.36) and (197.92,167.27) .. (197.92,172.09) -- (197.92,198.27) .. controls (197.92,203.09) and (194.01,207) .. (189.19,207) -- (162.18,207) .. controls (157.36,207) and (153.45,203.09) .. (153.45,198.27) -- cycle ;
\draw [line width=0.75]    (24.75,189.3) -- (64.16,188.79) ;
\draw [shift={(66.16,188.76)}, rotate = 179.26] [fill={rgb, 255:red, 0; green, 0; blue, 0 }  ][line width=0.08]  [draw opacity=0] (7.2,-1.8) -- (0,0) -- (7.2,1.8) -- cycle    ;
\draw [line width=0.75]    (21.76,100.23) -- (61.17,99.73) ;
\draw [shift={(63.17,99.7)}, rotate = 179.26] [fill={rgb, 255:red, 0; green, 0; blue, 0 }  ][line width=0.08]  [draw opacity=0] (7.2,-1.8) -- (0,0) -- (7.2,1.8) -- cycle    ;
\draw [line width=0.75]    (319,40) -- (352.16,97.11) ;
\draw [shift={(353.08,98.89)}, rotate = 242.83] [fill={rgb, 255:red, 0; green, 0; blue, 0 }  ][line width=0.08]  [draw opacity=0] (7.2,-1.8) -- (0,0) -- (7.2,1.8) -- cycle    ;
\draw  [color={rgb, 255:red, 21; green, 20; blue, 20 }  ,draw opacity=1 ][fill={rgb, 255:red, 230; green, 230; blue, 230  }  ,fill opacity=1 ] (150.45,25.01) .. controls (150.45,20.19) and (154.36,16.28) .. (159.18,16.28) -- (186.19,16.28) .. controls (191.01,16.28) and (194.92,20.19) .. (194.92,25.01) -- (194.92,51.19) .. controls (194.92,56.01) and (191.01,59.92) .. (186.19,59.92) -- (159.18,59.92) .. controls (154.36,59.92) and (150.45,56.01) .. (150.45,51.19) -- cycle ;
\draw [line width=0.75]    (22.74,42.22) -- (62.15,41.71) ;
\draw [shift={(64.15,41.68)}, rotate = 179.26] [fill={rgb, 255:red, 0; green, 0; blue, 0 }  ][line width=0.08]  [draw opacity=0] (7.2,-1.8) -- (0,0) -- (7.2,1.8) -- cycle    ;
\draw  [color={rgb, 255:red, 21; green, 20; blue, 20 }  ,draw opacity=1 ][fill={rgb, 255:red, 230; green, 230; blue, 230 }  ,fill opacity=1 ] (392.13,95.97) .. controls (392.13,91.15) and (396.03,87.24) .. (400.85,87.24) -- (427.86,87.24) .. controls (432.68,87.24) and (436.59,91.15) .. (436.59,95.97) -- (436.59,122.16) .. controls (436.59,126.98) and (432.68,130.89) .. (427.86,130.89) -- (400.85,130.89) .. controls (396.03,130.89) and (392.13,126.98) .. (392.13,122.16) -- cycle ;
\draw  [color={rgb, 255:red, 21; green, 20; blue, 20 }  ,draw opacity=1 ][fill={rgb, 255:red, 230; green, 230; blue, 230  }  ,fill opacity=1] (468.37,94.97) .. controls (468.37,90.15) and (472.28,86.24) .. (477.1,86.24) -- (504.11,86.24) .. controls (508.93,86.24) and (512.84,90.15) .. (512.84,94.97) -- (512.84,121.16) .. controls (512.84,125.98) and (508.93,129.89) .. (504.11,129.89) -- (477.1,129.89) .. controls (472.28,129.89) and (468.37,125.98) .. (468.37,121.16) -- cycle ;
\draw [line width=0.75]    (437.27,108.99) -- (465.5,109) ;
\draw [shift={(467.5,109)}, rotate = 180.02] [fill={rgb, 255:red, 0; green, 0; blue, 0 }  ][line width=0.08]  [draw opacity=0] (7.2,-1.8) -- (0,0) -- (7.2,1.8) -- cycle    ;
\draw  [color={rgb, 255:red, 21; green, 20; blue, 20 }  ,draw opacity=1 ][fill={rgb, 255:red, 230; green, 230; blue, 230 }  ,fill opacity=1] (238.46,85.03) .. controls (238.46,80.2) and (242.37,76.3) .. (247.19,76.3) -- (274.2,76.3) .. controls (279.02,76.3) and (282.93,80.2) .. (282.93,85.03) -- (282.93,111.21) .. controls (282.93,116.03) and (279.02,119.94) .. (274.2,119.94) -- (247.19,119.94) .. controls (242.37,119.94) and (238.46,116.03) .. (238.46,111.21) -- cycle ;
\draw  [color={rgb, 255:red, 21; green, 20; blue, 20 }  ,draw opacity=1 ][fill={rgb, 255:red, 230; green, 230; blue, 230  }  ,fill opacity=1 ] (241.45,173.09) .. controls (241.45,168.27) and (245.36,164.36) .. (250.18,164.36) -- (277.19,164.36) .. controls (282.01,164.36) and (285.92,168.27) .. (285.92,173.09) -- (285.92,199.27) .. controls (285.92,204.09) and (282.01,208) .. (277.19,208) -- (250.18,208) .. controls (245.36,208) and (241.45,204.09) .. (241.45,199.27) -- cycle ;
\draw  [color={rgb, 255:red, 21; green, 20; blue, 20 }  ,draw opacity=1 ][fill={rgb, 255:red, 230; green, 230; blue, 230  }  ,fill opacity=1 ] (238.45,26.01) .. controls (238.45,21.19) and (242.36,17.28) .. (247.18,17.28) -- (274.19,17.28) .. controls (279.01,17.28) and (282.92,21.19) .. (282.92,26.01) -- (282.92,52.19) .. controls (282.92,57.01) and (279.01,60.92) .. (274.19,60.92) -- (247.18,60.92) .. controls (242.36,60.92) and (238.45,57.01) .. (238.45,52.19) -- cycle ;
\draw [line width=0.75]    (286.35,187.11) -- (319,187.01) ;
\draw [line width=0.75]    (284.35,98.11) -- (319,99) ;
\draw [line width=0.75]    (283.35,39.11) -- (319,40) ;
\draw  [color={rgb, 255:red, 21; green, 20; blue, 20 }  ,draw opacity=1 ][fill={rgb, 255:red, 230; green, 230; blue, 230  }  ,fill opacity=1 ] (63.46,85.09) .. controls (63.46,80.27) and (67.37,76.36) .. (72.19,76.36) -- (99.2,76.36) .. controls (104.02,76.36) and (107.93,80.27) .. (107.93,85.09) -- (107.93,111.27) .. controls (107.93,116.09) and (104.02,120) .. (99.2,120) -- (72.19,120) .. controls (67.37,120) and (63.46,116.09) .. (63.46,111.27) -- cycle ;
\draw  [color={rgb, 255:red, 21; green, 20; blue, 20 }  ,draw opacity=1 ][fill={rgb, 255:red, 230; green, 230; blue, 230  }  ,fill opacity=1 ] (66.45,173.15) .. controls (66.45,168.33) and (70.36,164.42) .. (75.18,164.42) -- (102.19,164.42) .. controls (107.01,164.42) and (110.92,168.33) .. (110.92,173.15) -- (110.92,199.33) .. controls (110.92,204.15) and (107.01,208.06) .. (102.19,208.06) -- (75.18,208.06) .. controls (70.36,208.06) and (66.45,204.15) .. (66.45,199.33) -- cycle ;
\draw  [color={rgb, 255:red, 21; green, 20; blue, 20 }  ,draw opacity=1 ][fill={rgb, 255:red, 230; green, 230; blue, 230  }  ,fill opacity=1 ] (63.45,26.07) .. controls (63.45,21.25) and (67.36,17.34) .. (72.18,17.34) -- (99.19,17.34) .. controls (104.01,17.34) and (107.92,21.25) .. (107.92,26.07) -- (107.92,52.25) .. controls (107.92,57.07) and (104.01,60.98) .. (99.19,60.98) -- (72.18,60.98) .. controls (67.36,60.98) and (63.45,57.07) .. (63.45,52.25) -- cycle ;
\draw [line width=0.75]    (110.75,188.86) -- (150.16,188.35) ;
\draw [shift={(152.16,188.32)}, rotate = 179.26] [fill={rgb, 255:red, 0; green, 0; blue, 0 }  ][line width=0.08]  [draw opacity=0] (7.2,-1.8) -- (0,0) -- (7.2,1.8) -- cycle    ;
\draw [line width=0.75]    (107.76,99.8) -- (147.17,99.29) ;
\draw [shift={(149.17,99.26)}, rotate = 179.26] [fill={rgb, 255:red, 0; green, 0; blue, 0 }  ][line width=0.08]  [draw opacity=0] (7.2,-1.8) -- (0,0) -- (7.2,1.8) -- cycle    ;
\draw [line width=0.75]    (108.74,41.78) -- (148.15,41.27) ;
\draw [shift={(150.15,41.24)}, rotate = 179.26] [fill={rgb, 255:red, 0; green, 0; blue, 0 }  ][line width=0.08]  [draw opacity=0] (7.2,-1.8) -- (0,0) -- (7.2,1.8) -- cycle    ;
\draw [line width=0.75]    (198.75,188.3) -- (238.16,187.79) ;
\draw [shift={(240.16,187.76)}, rotate = 179.26] [fill={rgb, 255:red, 0; green, 0; blue, 0 }  ][line width=0.08]  [draw opacity=0] (7.2,-1.8) -- (0,0) -- (7.2,1.8) -- cycle    ;
\draw [line width=0.75]    (195.76,99.23) -- (235.17,98.73) ;
\draw [shift={(237.17,98.7)}, rotate = 179.26] [fill={rgb, 255:red, 0; green, 0; blue, 0 }  ][line width=0.08]  [draw opacity=0] (7.2,-1.8) -- (0,0) -- (7.2,1.8) -- cycle    ;
\draw [line width=0.75]    (196.74,41.22) -- (236.15,40.71) ;
\draw [shift={(238.15,40.68)}, rotate = 179.26] [fill={rgb, 255:red, 0; green, 0; blue, 0 }  ][line width=0.08]  [draw opacity=0] (7.2,-1.8) -- (0,0) -- (7.2,1.8) -- cycle    ;

\draw (622.6,104.09) node  [font=\footnotesize] [align=left] {$\hat{g}^n$};
\draw (41.14,86.25) node  [font=\normalsize] [align=left] {$g_{2}^n$};
\draw (39.92,143.83) node  [font=\Large] [align=left] {$\vdots$};
\draw (378.91,96.13) node  [font=\footnotesize] [align=left] {$\bm{y}^n$};
\draw (360.36,31.69) node  [font=\normalsize] [align=left] {$\bm{z}^n$};
\draw (157.02,90.63) node [anchor=north west][inner sep=0.75pt]    {$\mathscr{E}_q(\cdot)$};
\draw (44.13,177.5) node  [font=\normalsize] [align=left] {$g_{K}^n$};
\draw (408,105) node [anchor=north west][inner sep=0.75pt]    {$\bm{u}^n$};
\draw (549.37,101.37) node [anchor=north west][inner sep=0.75pt]  {$\mathscr{D}_q(\cdot)$};
\draw (41.13,28.23) node  [font=\normalsize] [align=left] {$g_{1}^n$};
\draw (158.01,29.33) node [anchor=north west][inner sep=0.75pt]    {$\mathscr{E}_q(\cdot)$};
\draw (530.14,99.32) node  [font=\normalsize] [align=left] {$r^n$};
\draw (475,101.4) node [anchor=north west][inner sep=0.75pt]    {$\psi(\cdot)$};
\draw (250,29.41) node [anchor=north west][inner sep=0.75pt]    {$\varphi(\cdot)$};
\draw (250,87.43) node [anchor=north west][inner sep=0.75pt]    {$\varphi(\cdot)$};
\draw (250,176.49) node [anchor=north west][inner sep=0.75pt]    {$\varphi(\cdot)$};
\draw (73,29.47) node [anchor=north west][inner sep=0.75pt]    {$\mathcal{Q}_q(\cdot)$};
\draw (73,87.49) node [anchor=north west][inner sep=0.75pt]    {$\mathcal{Q}_q(\cdot)$};
\draw (73,176.55) node [anchor=north west][inner sep=0.75pt]    {$\mathcal{Q}_q(\cdot)$};
\draw (215.14,82.81) node  [font=\normalsize] [align=left] {$x_{2}^n$};
\draw (125.92,143.39) node  [font=\Large] [align=left] {$\vdots$};
\draw (216.13,173.06) node  [font=\normalsize] [align=left] {$x_{K}^n$};
\draw (216.13,27.79) node  [font=\normalsize] [align=left] {$x_{1}^n$};
\draw (310.13,26.79) node  [font=\normalsize] [align=left] {$s_{1}^n\times \bm{h}_{1}^n$};

\draw (310,84.81) node  [font=\normalsize] [align=left] {$s_{2}^n\times \bm{h}_{2}^n$};
\draw (313,194.81) node  [font=\normalsize] [align=left] {$s_{K}^n\times \bm{h}_{K}^n$};
\draw (213.92,142.83) node  [font=\Large] [align=left] {$ \vdots $};
\draw (129.13,24.23) node  [font=\normalsize] [align=left] {$\tilde{g}_{1}^n$};
\draw (157.45,177.49) node [anchor=north west][inner sep=0.75pt]   {$\mathscr{E}_q( \cdot )$};
\draw (128.13,80.23) node  [font=\normalsize] [align=left] {$ \tilde{g}_{2}^n$};
\draw (131.13,173.23) node  [font=\normalsize] [align=left] {$ \tilde{g}_{K}^n$};
\draw (453.91,96.13) node  [font=\normalsize] [align=left] {$\hat{s}^n$};

\end{tikzpicture}

}
 
    \caption{Block diagram illustrating the communication model for \ac{FEEL} at the $n$-th communication subchannel. The gradients  $g_{1}^n$, $g_{2}^n, \ldots, g_{K}^n$ are first quantized using operator $\mathcal{Q}_q(\cdot)$, and then go through an encoder, $\mathscr{E}_q(\cdot)$, and the pre-processing function, $\varphi$. Then, all edge devices transmit the modulated signals $s_{1}^n$, $s_{2}^n, \ldots s_{K}^n$  over the \ac{MAC} resulting in the received vector $\bm{y}^n$, which is degraded by the noise $\bm{z}^n$ and the wireless channel effects $\bm{h}^n$. The received signal by $N_r$ antennas is $\bm{y}^n$, which undergoes a receiver beamforming vector, $\bm{u}^n$, at the \ac{ES}. Then, the resultant signal is passed through the post-processing function $\psi$ to obtain $n$-th element of the received vector $\bm{r}$, i.e., $r^n$. Finally, $r^n$  is decoded by the decoder $\mathscr{D}_q(\cdot)$ to yield the estimated function $\hat{g}^n$.  }
    \label{fig:SumCompFEEL}
\end{figure*}

%% file: fig/Fig_Concenrtare.tex
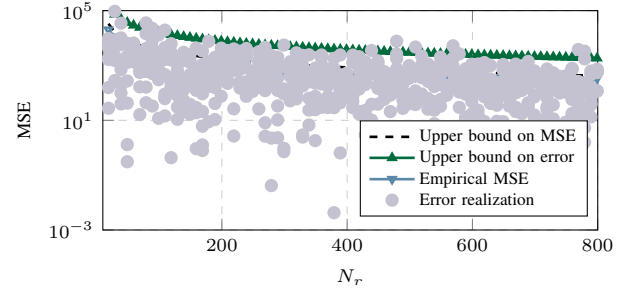
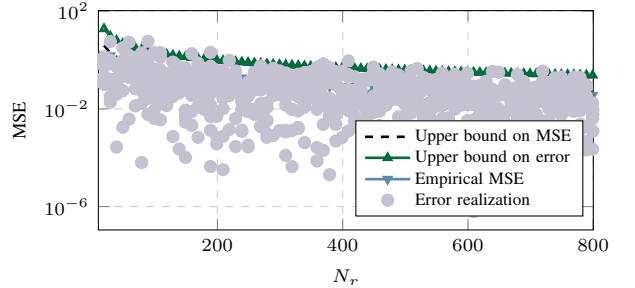
\begin{figure}[!t]
\centering
\subfigure[Estimation of $s^n$]{\label{fig:ConcentrateS(a)}
    \begin{tikzpicture} 
    \begin{axis}[
        xlabel={ $N_r$},
        ylabel={MSE},
        label style={font=\scriptsize},
        legend cell align={left},
        tick label style={font=\scriptsize} , 
        width=0.45\textwidth,
        height=4.5cm,
        xmin=10, xmax=800,
        ymin=1e-3, ymax=10^5,
        axis on top,
        ymode = log,
       legend style={nodes={scale=0.65, transform shape}, at={(0.98,0.5)}}, 
        ymajorgrids=true,
        xmajorgrids=true,
        grid style=dashed,
        grid=both,
        grid style={line width=.1pt, draw=gray!15},
        major grid style={line width=.2pt,draw=gray!40},
    ]
  \addplot[
        color=black,
        dashed,
        line width=1pt,
        ]
    table[x=Antenna,y=MSETE]
    {Data/MSENumber.dat};
    \addlegendentry{Upper bound on MSE}
     \addplot[
        color=cadmiumgreen,
        mark=triangle,
        line width=1pt,
        mark size=1.5pt,
        ]
    table[x=Antenna,y=MSET1]
    {Data/MSENumber.dat};
     \addlegendentry{Upper bound on error}
     \addplot[
    color=airforceblue,
    mark=triangle,
    mark options = {rotate = 180},
    line width=1pt,
    mark size=1.5pt,
    ]
table[x=Antenna,y=MSE]
{Data/MSENumber.dat};
 \addlegendentry{ Empirical MSE}
 \addplot[color=lavendergray, 
        only marks,
         fill opacity=0.1,
        draw opacity=0,
        line width=1pt,
        ]
table[x=Antenna,y=P0]
{Data/MSENEmprical.dat};
 \addlegendentry{Error realization}
 \addplot[color=lavendergray,
        only marks,
        fill opacity=0.1,
        draw opacity=0,
        line width=1pt,
        ]
table[x=Antenna,y=P1]
{Data/MSENEmprical.dat};
 \addplot[color=lavendergray, 
        only marks,
        fill opacity=0.1,
        draw opacity=0,
        line width=1pt,
        ]
table[x=Antenna,y=P2]
{Data/MSENEmprical.dat};
 \addplot[color=lavendergray, 
        only marks,
        fill opacity=0.1,
        draw opacity=0,
        line width=1pt,
        ]
table[x=Antenna,y=P3]
{Data/MSENEmprical.dat};
 \addplot[color=lavendergray, 
        only marks,
        fill opacity=0.1,
        draw opacity=0,
        line width=1pt,
        ]
table[x=Antenna,y=P4]
{Data/MSENEmprical.dat};
 \addplot[color=lavendergray, 
        only marks,
        fill opacity=0.1,
        draw opacity=0,
        line width=1pt,
        ]
table[x=Antenna,y=P5]
{Data/MSENEmprical.dat};
 \addplot[color=lavendergray, 
        only marks,
        fill opacity=0.1,
        draw opacity=0,
        line width=1pt,
        ]
table[x=Antenna,y=P6]
{Data/MSENEmprical.dat};
 \addplot[color=lavendergray, 
        only marks,
        fill opacity=0.1,
        draw opacity=0,
        line width=1pt,
        ]
table[x=Antenna,y=P7]
{Data/MSENEmprical.dat};
 \addplot[color=lavendergray, 
        only marks,
        fill opacity=0.1,
        draw opacity=0,
        line width=1pt,
        ]
table[x=Antenna,y=P8]
{Data/MSENEmprical.dat};
    \end{axis}
\end{tikzpicture}}
\subfigure[Estimation of $\bm{g}$]{\label{fig:ConcentrateS(b)}
    \begin{tikzpicture} 
    \begin{axis}[
        xlabel={ $N_r$},
        ylabel={MSE},
        label style={font=\scriptsize},
        legend cell align={left},
        tick label style={font=\scriptsize} , 
        width=0.45\textwidth,
        height=4.5cm,
        xmin=10, xmax=800,
        axis on top,
        ymode = log,
       legend style={nodes={scale=0.65, transform shape}, at={(0.98,0.5)}}, 
        ymajorgrids=true,
        xmajorgrids=true,
        grid style=dashed,
        grid=both,
        grid style={line width=.1pt, draw=gray!15},
        major grid style={line width=.2pt,draw=gray!40},
    ]
  \addplot[
        color=black,
        dashed,
        line width=1pt,
        ]
    table[x=Antenna,y=MSETE]
    {Data/MSENumber2.dat};
    \addlegendentry{Upper bound on MSE}
     \addplot[
        color=cadmiumgreen,
        mark=triangle,
        line width=1pt,
        mark size=1.5pt,
        ]
    table[x=Antenna,y=MSET1]
    {Data/MSENumber2.dat};
     \addlegendentry{Upper bound on error}
     \addplot[
    color=airforceblue,
    mark=triangle,
    mark options = {rotate = 180},
    line width=1pt,
    mark size=1.5pt,
    ]
table[x=Antenna,y=MSE]
{Data/MSENumber2.dat};
 \addlegendentry{Empirical MSE }
 \addplot[
        color=lavendergray, 
        only marks,
        fill opacity=0.1,
        draw opacity=0,
        line width=1pt,
        ]
table[x=Antenna,y=P0]
{Data/MSENEmprical2.dat};
\addlegendentry{Error realization}
 \addplot[color=lavendergray, 
        only marks,
        fill opacity=0.1,
        draw opacity=0,
        line width=1pt,
        ]
table[x=Antenna,y=P1]
{Data/MSENEmprical2.dat};
 \addplot[
        color=lavendergray,
        only marks,
        fill opacity=0.1,
        draw opacity=0,
        line width=1pt,
        ]
table[x=Antenna,y=P2]
{Data/MSENEmprical2.dat};
 \addplot[
        color=lavendergray, 
        only marks,
        fill opacity=0.1,
        draw opacity=0,
        line width=1pt,
        ]
table[x=Antenna,y=P3]
{Data/MSENEmprical2.dat};
 \addplot[
        color=lavendergray, 
        only marks,
        fill opacity=0.1,
        draw opacity=0,
        line width=1pt,
        ]
table[x=Antenna,y=P4]
{Data/MSENEmprical2.dat};
 \addplot[
        color=lavendergray, 
        only marks,
        fill opacity=0.1,
        draw opacity=0,
        line width=1pt,
        ]
table[x=Antenna,y=P5]
{Data/MSENEmprical2.dat};
 \addplot[
        color=lavendergray, 
        only marks,
        fill opacity=0.1,
        draw opacity=0,
        line width=1pt,
        ]
table[x=Antenna,y=P6]
{Data/MSENEmprical2.dat};
 \addplot[
        color=lavendergray,
        only marks,
        fill opacity=0.1,
        draw opacity=0,
        line width=1pt,
        ]
table[x=Antenna,y=P7]
{Data/MSENEmprical2.dat};
 \addplot[
        color=lavendergray,
        only marks,
        fill opacity=0.1,
        draw opacity=0,
        line width=1pt,
        ]
table[x=Antenna,y=P8]
{Data/MSENEmprical2.dat};
    \end{axis}
\end{tikzpicture}
}
  \caption{Monte Carlo numerical evaluation of  the summation function, for $10$ trials versus the analytical results from Theorem~\ref{th:NRerror} for $\delta =0.01$ over different numbers of antennas, $N_r$. The channel coefficients and channel noise  generated by $N(\bm{0}, \sigma_h\bm{I}_{N_r})$ and $\mathcal{CN}(\bm{0}, \sigma_z\bm{I}_{N_r})$, respectively, are $\sigma_h = \sigma_z=1$.   Figure~\ref{fig:ConcentrateS(a)} shows the empirical \ac{MSE} of $\hat{s}^n$, analytical upper bound on the error in \eqref{eq:epsilonupp}, and expected value in \eqref{eq:Expecepsilonupp} form Theorem~\ref{th:NRerror}, for $K=200$ edge devices.
   Figure~\ref{fig:ConcentrateS(b)} shows the empirical and analytical upper bound on the \ac{MSE} of gradient $\hat{\bm{g}}$ from Proposition~\ref{cor:norm2}, for $K=20$ edge devices whose elements of their gradients, $\bm{g}_k$, generated uniformly at random from $\mathcal{U}[-2,2]$.  
    }
  \label{fig:ConcentrateS}
\end{figure}

%% file: fig/Fig_Gradient.tex
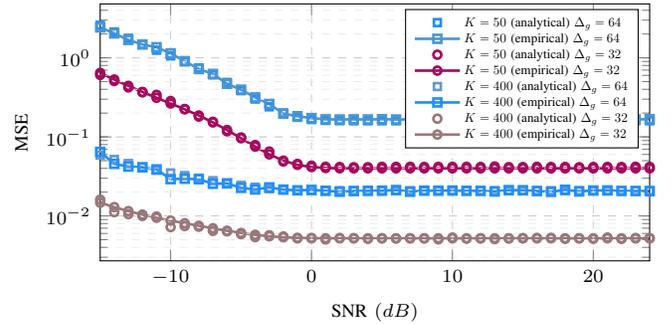
\begin{figure}
\centering
    \begin{tikzpicture} 
    \begin{axis}[
        xlabel={SNR $(dB)$},
        ylabel={MSE},
        label style={font=\scriptsize},
        legend cell align={left},
        tick label style={font=\scriptsize} , 
        width=0.49\textwidth,
        height=5cm,
        xmin=-15, xmax=24,
        axis on top,
        ymode = log,
        legend style={nodes={scale=0.49, transform shape}, at={(0.98,0.98)}}, 
        ymajorgrids=true,
        xmajorgrids=true,
        grid style=dashed,
        grid=both,
        grid style={line width=.1pt, draw=gray!15},
        major grid style={line width=.2pt,draw=gray!40},
    ]
    \addplot[only marks,
        color=dodgerblue,
        mark=square,
        line width=1pt,
        mark size=1.3pt,
        ]
    table[x=SNR,y=MSES1]
    {Data/MSEGrad.dat};
    \addplot[
        color=celestialblue,
        mark=square,
        line width=1pt,
        mark size=1.5pt,
        ]
    table[x=SNR,y=MSET1]
    {Data/MSEGrad.dat};
         \addplot[only marks,
        color=jazzberryjam,
        mark=o,
        line width=1pt,
        mark size=1.5pt,
        ]
    table[x=SNR,y=MSES2]
    {Data/MSEGrad.dat};
    \addplot[
        color=jazzberryjam,
        mark=o,
        line width=1pt,
        mark size=1.5pt,
        ]
    table[x=SNR,y=MSET2]
    {Data/MSEGrad.dat};
    \addplot[only marks,
        color=bluegray,
        mark=square,
        line width=1pt,
        mark size=1.3pt,
        ]
    table[x=SNR,y=MSES1]
    {Data/MSEGrad400.dat};
    \addplot[
        color=dodgerblue,
        mark=square,
        line width=1pt,
        mark size=1.5pt,
        ]
    table[x=SNR,y=MSET1]
    {Data/MSEGrad400.dat};
         \addplot[only marks,
        color=bazaar,
        mark=o,
        line width=1pt,
        mark size=1.5pt,
        ]
    table[x=SNR,y=MSES2]
    {Data/MSEGrad400.dat};
    \addplot[
        color=bazaar,
        mark=o,
        line width=1pt,
        mark size=1.5pt,
        ]
    table[x=SNR,y=MSET2]
    {Data/MSEGrad400.dat};
    
    \legend{$K=50$  (analytical) $\Delta_g = 64$, $K=50$ (empirical) $\Delta_g = 64$,$K=50$ (analytical) $\Delta_g = 32$,$K=50$ (empirical) $\Delta_g = 32$, $K=400$  (analytical) $\Delta_g = 64$, $K=400$ (empirical) $\Delta_g = 64$,$K=400$ (analytical) $\Delta_g = 32$,$K=400$ (empirical) $\Delta_g = 32$  };
    \end{axis}
\end{tikzpicture}
  \caption{  \change Monte Carlo numerical evaluation of the average gradient estimation in \eqref{eq:MSEGradiant} with true gradient, for $100$ trials versus the analytical results from Proposition~\ref{Pr:MSE}. Here, we consider $q = 64$ and $N = 100$ for two cases of $K=50$ and $K=400$ edge devices.  The gradients $\bm{g}_k$ generated uniformly at random from $\mathcal{U}[0,64]$ and $\mathcal{U}[0,32]$.   }
  \label{fig:Gradient}
\end{figure}

%% file: fig/Fig_MNIST.tex
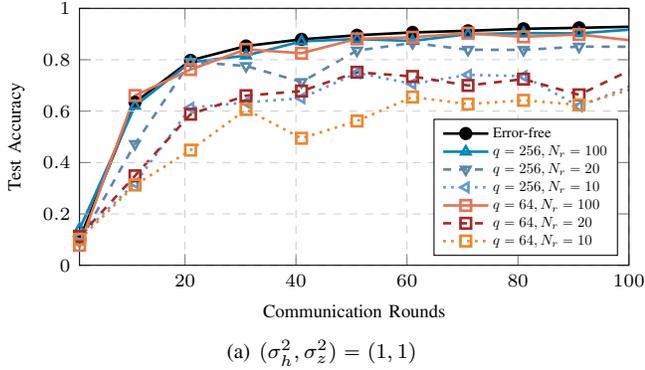
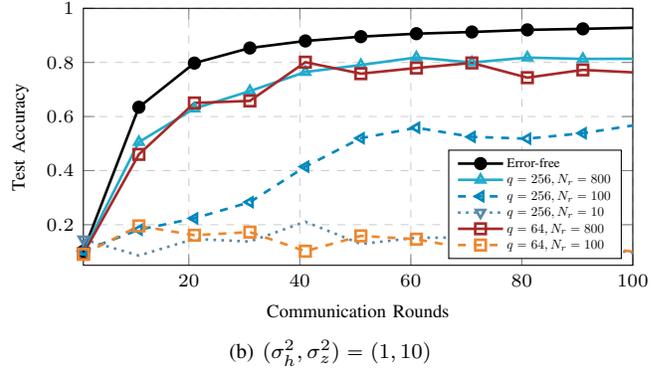
\begin{figure*}
\centering
\subfigure[$(\sigma_h^2,\sigma_z^2) = (1,1)$]{\label{fig:MNIST(a)}
    \begin{tikzpicture} 
    \begin{axis}[
        xlabel={ Communication Rounds},
        ylabel={Test Accuracy},
        label style={font=\scriptsize},
        legend cell align={left},
        tick label style={font=\scriptsize} , 
        width=0.49\textwidth,
        height=5cm,
        xmin=1, xmax=100,
        ymin=0, ymax=1,
        axis on top,
        restrict y to domain=0:1,
       legend style={nodes={scale=0.55, transform shape}, at={(0.98,0.57)}}, 
        each nth point=10, filter discard warning=false, unbounded coords=discard, 
        ymajorgrids=true,
        xmajorgrids=true,
        grid style=dashed,
        grid=both,
        grid style={line width=.1pt, draw=gray!15},
        major grid style={line width=.2pt,draw=gray!40},
    ]
    \addlegendentry{Error-free}
     \addplot[
        color=black,
        mark= *,
        line width=1pt,
        ]
    table[x=Iter,y=AccQ1]
    {Data/MNISTIDEAL.dat};
\addlegendentry{$q=256, N_r = 100$}
\addplot[
    color=blue(ncs),
    mark=triangle,
    line width=1pt,
    mark size=2pt,
    ]
table[x=Iter,y=AccQ1]
    {Data/MNISTNt100q8.dat};
     \addlegendentry{$q=256, N_r = 20$}
     \addplot[
    color=airforceblue,
    dashed,
    mark=triangle,
    mark options = {solid, rotate = 180},
    line width=1pt,
    mark size=2pt,
    ]
table[x=Iter,y=AccQ1]
    {Data/MNIST.dat};
     \addlegendentry{$q=256, N_r = 10$}
     \addplot[
    color=bluegray,
    dotted,
    mark=triangle,
    mark options = {solid, rotate = 90},
    line width=1pt,
    mark size=2pt,
    ]
table[x=Iter,y=AccQ1]
    {Data/MNISTNt10q8.dat};
\addlegendentry{$q=64, N_r = 100$}
\addplot[
    color=burntsienna,
    mark=square,
    line width=1pt,
    mark size=2pt,
    ]
table[x=Iter,y=AccQ1]
    {Data/MNISTq4Nr100z1.dat};   
\addlegendentry{$q=64, N_r = 20$}
\addplot[
    color=brown(web),
    dashed,
    mark options={solid},
    mark=square,
    line width=1pt,
    mark size=2pt,
    ]
table[x=Iter,y=AccQ1]
    {Data/MNISTNr20q4z1.dat}; 
\addlegendentry{$q=64, N_r = 10$}
\addplot[
    color=cadmiumorange,
    dotted,
    mark options={solid},
    mark=square,
    line width=1pt,
    mark size=2pt,
    ]
table[x=Iter,y=AccQ1]
    {Data/MNISTq4Nr10z1.dat};
    \end{axis}
\end{tikzpicture}}
\subfigure[$(\sigma_h^2,\sigma_z^2) = (1,10)$]{\label{fig:MNIST(b)}
    \begin{tikzpicture} 
    \begin{axis}[
        xlabel={Communication Rounds},
        ylabel={Test Accuracy},
        label style={font=\scriptsize},
        legend cell align={left},
        tick label style={font=\scriptsize} , 
        width=0.49\textwidth,
        height=5cm,
        xmin=1, xmax=100,
        ymin=0.05, ymax=1,
        restrict y to domain=0:1,
        each nth point=10, filter discard warning=false, unbounded coords=discard, 
       legend style={nodes={scale=0.5, transform shape}, at={(0.98,0.45)}}, 
        ymajorgrids=true,
        xmajorgrids=true,
        grid style=dashed,
        grid=both,
        grid style={line width=.1pt, draw=gray!15},
        major grid style={line width=.2pt,draw=gray!40},
    ]
    \addlegendentry{Error-free}
     \addplot[
        color=black,
        mark= *,
        mark repeat=1,
        line width=1pt,
        ]
    table[x=Iter,y=AccQ1]
    {Data/MNISTIDEAL.dat};
\addlegendentry{$q=256, N_r = 800$}
\addplot[
    color=ballblue,
    mark=triangle,
    line width=1pt,
    mark size=2pt,
    ]
table[x=Iter,y=AccQ1]
    {Data/MNISTNr1600z10.dat};
\addlegendentry{$q=256, N_r = 100$}
\addplot[
    color=blue(ncs),
    dashed,
    mark=triangle,
    mark options = {solid, rotate = 90},
    line width=1pt,
    mark size=2pt,
    ]
table[x=Iter,y=AccQ1]
    {Data/MNISTNt100z10.dat};
\addlegendentry{$q=256, N_r = 10$}
     \addplot[
    color=airforceblue,
    dotted,
    mark repeat=10,
    mark=triangle,
    mark options = {solid, rotate = 180},
    line width=1pt,
    mark size=2pt,
    ]
table[x=Iter,y=AccQ1]
    {Data/MNISTNt10z10.dat};
\addlegendentry{$q=64, N_r = 800$}
\addplot[
    color=brown(web),
    mark=square,
    line width=1pt,
    mark size=2pt,
    ]
table[x=Iter,y=AccQ1]
    {Data/MNISTq4Nr800z10.dat};
\addlegendentry{$q=64, N_r = 100$}
\addplot[
    color=cadmiumorange,
    dashed,
    mark=square,
    mark options = {solid},
    line width=1pt,
    mark size=2pt,
    ]
table[x=Iter,y=AccQ1]
    {Data/MNISTNr100q4z10.dat};
    \end{axis}
\end{tikzpicture}
}
  \caption{Accuracy of the MNIST task as a function of the communication rounds for $K=20$ edge devices and heterogeneous data distribution across edge devices. Figures \ref{fig:MNIST(a)}  and \ref{fig:MNIST(b)} show the accuracy of \ac{FEEL} versus number of communication rounds for two low variance of the noise, i.e., $\sigma_z^2 =1$ and the high variance of the noise, i.e., $\sigma_z^2 =10$, respectively. }
  \label{fig:MNIST}
   
\end{figure*}

%% file: fig/Fig_IIDMNIST.tex
\begin{figure*}[!t]
\centering
\subfigure[$(\sigma_h^2, \sigma_z^2) = (5,2)$]{\label{fig:IDDlearning}
    \begin{tikzpicture} 
    \begin{axis}[
        xlabel={Communication Rounds},
        ylabel={Test Accuracy},
        label style={font=\scriptsize},
        legend cell align={left},
        tick label style={font=\scriptsize} , 
        width=0.45\textwidth,
        height=5cm,
        xmin=1, xmax=100,
        ymin=0.2, ymax=1,
        restrict y to domain=0:1,
        each nth point=5, filter discard warning=false, unbounded coords=discard, 
       legend style={nodes={scale=0.75, transform shape}, at={(0.98,0.65)}}, 
        ymajorgrids=true,
        xmajorgrids=true,
        grid style=dashed,
        grid=both,
        grid style={line width=.1pt, draw=gray!15},
        major grid style={line width=.2pt,draw=gray!40},
    ]
    \addlegendentry{Error-free}
     \addplot[
        color=black,
        mark= *,
        line width=1pt,
        ]
    table[x=Iter,y=AccQ1]
    {Data/IIDMNISTErrorfree.dat};
     \addlegendentry{Analog $N_r = 100$}
     \addplot[
    color=cadmiumorange,
    mark=square,
    mark options = {rotate = 180},
    line width=1pt,
    mark size=2pt,
    ]
table[x=Iter,y=AccQ1]
    {Data/IIDMNISTNr100z1.dat};
\addlegendentry{$q=256, N_r = 100$}
\addplot[
    color=blue(ncs),
    mark=triangle,
    line width=1pt,
    mark size=2pt,
    ]
table[x=Iter,y=AccQ1]
    {Data/IIDMNISTq8Nr100z10.dat};
\addlegendentry{Analog  $N_r = 400$}
\addplot[
    color=brown(web),
    mark=square,
    line width=1pt,
    mark size=2pt,
    ]
table[x=Iter,y=AccQ1]
    {Data/IIDMNISTNr400z10.dat};
\addlegendentry{$q=256, N_r = 400$}
\addplot[
    color=dodgerblue,
    mark=triangle,
    line width=1pt,
    mark size=2pt,
    ]
table[x=Iter,y=AccQ1]
    {Data/IIDMNISTq8Nr400z10.dat};
    \end{axis}
\end{tikzpicture}} \subfigure[$(\sigma_h^2, \sigma_z^2) = (2,2)$]{ \label{fig:IIDCIFAIR}
    \begin{tikzpicture} 
    \begin{axis}[
        xlabel={Communication Rounds},
        ylabel={Test Accuracy},
        label style={font=\scriptsize},
        legend cell align={left},
        tick label style={font=\scriptsize} , 
        width=0.45\textwidth,
        height=5cm,
        xmin=1, xmax=100,
        ymin=0.01, ymax=0.8,
        restrict y to domain=0:1,
        each nth point=5, filter discard warning=false, unbounded coords=discard, 
       legend style={nodes={scale=0.65, transform shape}, at={(0.98,0.55)}}, 
        ymajorgrids=true,
        xmajorgrids=true,
        grid style=dashed,
        grid=both,
        grid style={line width=.1pt, draw=gray!15},
        major grid style={line width=.2pt,draw=gray!40},
    ]
    \addlegendentry{Error-free}
     \addplot[
        color=black,
        mark= *,
        line width=1pt,
        ]
    table[x=Iter,y=AccQ1]
    {Data/CIFAIRIdeal.dat};
     \addlegendentry{Analog $N_r = 100$}
     \addplot[
    color=cadmiumorange,
    mark=square,
    mark options = {rotate = 180},
    line width=1pt,
    mark size=2pt,
    ]
table[x=Iter,y=AccQ1]
    {Data/CIFAIRAmiri100.dat};
\addlegendentry{$q=256, N_r = 100$}
\addplot[
    color=blue(ncs),
    mark=triangle,
    line width=1pt,
    mark size=2pt,
    ]
table[x=Iter,y=AccQ1]
    {Data/CIFAIRSaeed100.dat};
\addlegendentry{Analog  $N_r = 400$}
\addplot[
    color=brown(web),
    mark=square,
    line width=1pt,
    mark size=2pt,
    ]
table[x=Iter,y=AccQ1]
    {Data/CIFAIRAmiri400.dat};
\addlegendentry{$q=256, N_r = 400$}
\addplot[
    color=dodgerblue,
    mark=triangle,
    line width=1pt,
    mark size=2pt,
    ]
table[x=Iter,y=AccQ1]
    {Data/CIFAIRSaeed400.dat};
    \end{axis}
\end{tikzpicture}
}
  \caption{ \change Performance comparison for homogeneous dataset distribution among the edge devices. The accuracy is depicted versus the communication rounds for the error-free baseline, digital \ac{FEEL}, and the analog \ac{FEEL} method proposed in \cite{amiri2021blind}.  The channel coefficients are generated randomly with variance $\sigma_h^2=5$ and $\sigma_h^2=2$,  for Figure~\ref{fig:IDDlearning} and \ref{fig:IIDCIFAIR}. Also, the noise follows a Normal distribution with variance $\sigma_z^2= 2$ and  $\sigma_z^2= 2$ for Figure~\ref{fig:IDDlearning} and \ref{fig:IIDCIFAIR}, respectively. }
  \label{fig:IIDMNIST}
   
\end{figure*}
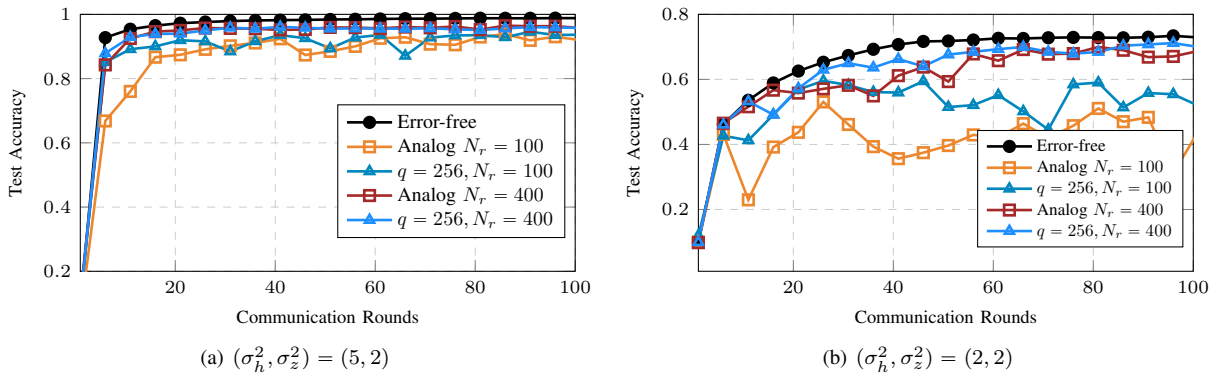

%% file: fig/Fig_Latency.tex
\begin{figure}[!t]
\centering

    \begin{tikzpicture} 
    \begin{axis}[
        xlabel={Edge devices($K$)},
        ylabel={Communication latency},
        label style={font=\scriptsize},
        tick label style={font=\scriptsize} , 
        width=0.45\textwidth,
        height=4.5cm,
        xmin=10, xmax=500,
         ymode = log,
       legend style={nodes={scale=0.65, transform shape}, at={(0.9,0.8)}}, 
        ymajorgrids=true,
        xmajorgrids=true,
        grid style=dashed,
        grid=both,
        grid style={line width=.1pt, draw=gray!15},
        major grid style={line width=.2pt,draw=gray!40},
    ]
\addlegendentry{OFDMA $q=256, N_r =400$}
    \addplot[
        color=black,
        mark=o,
        line width=1pt,
        mark size=2pt,
        ]
    table[x=device,y=OFDM400]
    {Data/latency.dat};
\addlegendentry{Analog, $ N_r = 100$}
    \addplot[
        color=cadmiumorange,
        mark= square,
        line width=1pt,
        mark size=2pt,
        ]
    table[x=device,y=Analog100]
    {Data/latency.dat};
\addlegendentry{Ours, $q=256, N_r = 100$}
     \addplot[
        color=blue(ncs),
        mark=triangle,
        line width=1pt,
        mark size=1.5pt,
        ]
    table[x=device,y=FLSumComp100]
    {Data/latency.dat};
\addlegendentry{Analog, $ N_r = 400$}
    \addplot[
        color=brown(web),
        mark= square,
        line width=1pt,
        mark size=2pt,
        ]
    table[x=device,y=Analog400]
    {Data/latency.dat};
    \addlegendentry{Ours, $q=256, N_r = 400$}
     \addplot[
        color=dodgerblue,
        mark=triangle,
        line width=1pt,
        mark size=1.5pt,
        ]
    table[x=device,y=FLSumComp400]
    {Data/latency.dat};
    \end{axis}
\end{tikzpicture}
  \caption{ {\change Comparison of latency reduction in FEEL: This figure illustrates the latency reduction achieved by ChannelComFed, analog FEEL, and OFDMA in federated learning tasks, considering a broadband communication setting characterized by a bandwidth of $B = 1$ kHz.}}
    \label{fig:latency}
\end{figure}
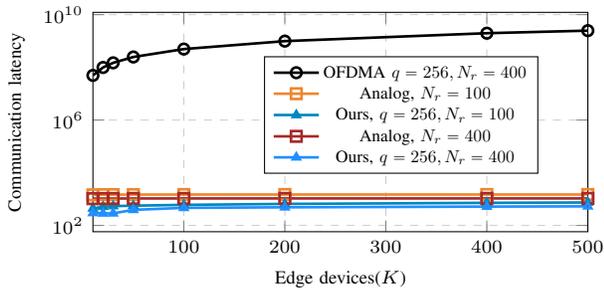